\documentclass[preprint,12pt]{elsarticle}

\newif\ifanonymous

\usepackage[utf8]{inputenc}
\usepackage[T1]{fontenc}
\usepackage{xcolor}
\usepackage{subcaption} %
\usepackage{rotating} %
\usepackage{calc}
\usepackage{tabularx}
\usepackage{adjustbox}
\usepackage[most]{tcolorbox}
\usepackage{longtable}
\usepackage{booktabs}

\newcommand{\navi}{\texttt{NAVI} }
\newcommand{\naviw}{\texttt{NAVI}}
\newcommand{\adi}[1]{#1}
\newcommand{\dk}[1]{#1}
\newcommand{\nuria}[1]{#1}

\usepackage{amssymb}
\usepackage{amsmath}

\journal{Computers in Human Behavior: Artificial Humans}

\usepackage[hidelinks]{hyperref}
\myfooterfont{\scriptsize\itshape}
\myfooter[C]{Published in \emph{Computers in Human Behavior: Artificial Humans}, \href{https://doi.org/10.1016/j.chbah.2026.100362}{doi:10.1016/j.chbah.2026.100362}}

\begin{document}

\begin{frontmatter}

\title{Mind the Style: Impact of Communication Style on Human-Chatbot Interaction}

\ifanonymous
    \author{Anonymous Authors}
    \address{Anonymous Institutions}
\else
    \author[EA,CTU]{Erik Derner}
    \author[USB]{Dalibor Ku\v{c}era}
    \author[EA]{Aditya Gulati}
    \author[UU]{Robert A. Bagheri}
    \author[EA]{Nuria Oliver}
    
    \affiliation[EA]{organization={ELLIS Alicante},%
                city={Alicante},
                country={Spain}}
    
    \affiliation[CTU]{organization={Czech Technical University in Prague},%
                city={Prague},
                country={Czech Republic}}
    
    \affiliation[USB]{organization={University of South Bohemia},%
                city={České Budějovice},
                country={Czech Republic}}
    
    \affiliation[UU]{organization={Utrecht University},%
                city={Utrecht},
                country={The Netherlands}}
\fi
\begin{abstract}
Conversational agents increasingly mediate everyday digital interactions, yet the effects of their communication style on user experience and task success remain \dk{insufficiently understood.} Addressing this gap, we \dk{report} a between-subject user study \dk{in which} participants \dk{interacted} with one of two versions of a chatbot called \naviw\dk{,} which \dk{assisted them} in an interactive map-based 2D navigation task. The two chatbot versions \dk{were designed to} differ \dk{primarily} in communication style: one \dk{used a} friendly and \dk{supportive tone,} while the other \dk{used a} direct and \dk{task-focused tone. We also included a control condition where participants did not interact with a chatbot but received the step-by-step navigation instructions.} The friendly chatbot significantly \dk{increased} users' \dk{communication} satisfaction \dk{and was} associated with higher task success \dk{than the direct chatbot. However, participants in the control condition achieved the highest task success overall, suggesting that chatbot interaction may introduce overhead in tasks that can be completed effectively using straightforward instructions. We did not find significant evidence that gender moderated the effects of communication style, although exploratory gender-stratified analyses suggested patterns that warrant further investigation. Finally, we found limited} evidence of \dk{global} linguistic \dk{accommodation, with only selective feature-level alignment.} These findings \dk{suggest} that \dk{chatbot communication style influences users' perceptions of conversational agents and may improve performance relative to less supportive chatbot designs, but the overall value of chatbot interaction depends on the task context. The study highlights the need for task-sensitive, transparent and carefully evaluated communication-style choices in conversational-agent design.}
\end{abstract}


\end{frontmatter}

\section{Introduction}

Conversational agents are increasingly used for task-oriented assistance, including in navigation, learning and decision-making scenarios \cite{Chowa2025, Yi2025}. Their effectiveness depends not only on informational accuracy but also on \nuria{their communication style}. Stylistic cues such as tone, politeness, warmth \dk{and brevity can shape users satisfaction, engagement and their judgment of the agent's competence}, making communication style an important design variable in human–AI interaction \cite{Go2019, Feine2019,cai2024communication,ding2024}.

\dk{In this paper, we use the term communication style in an operational sense. We do not treat ``friendly'' and ``direct'' as universal interpersonal types, nor as stable personalities of the conversational agent. Rather, they refer to two prompt-engineered modes of expressing the same navigation guidance. The friendly style operationalizes warmth, supportiveness, politeness and encouragement, whereas the direct style operationalizes brevity, task focus and minimal social framing. This distinction is grounded in work on politeness and relational communication \cite{brown1987politeness}, social responses to media \cite{reeves1996media}, and socio-cognitive models of warmth and competence \cite{fiske2002model}. It is also practically relevant because contemporary Large Language Model (LLM) based chatbots can be instructed to vary style while keeping their task goal constant.}

Social Response Theory proposes that people apply social norms to technologies that display human-like cues, even when they know that the system is artificial \cite{reeves1996media,nass2000machines}. Thus, linguistic style functions as a social signal that shapes the users' affective reactions and expectations. Communication Accommodation Theory complements this view by explaining how interlocutors adjust their linguistic behavior (\emph{e.g.,} lexical choices, pronouns, affective expressions, directness or temporal/numerical references) to manage affiliation, reduce social distance or express identity \cite{giles2023communication, giles2013communication}. Expectancy-based models additionally suggest that \nuria{communication style} \dk{matters because it} can \dk{confirm or violate the users' assumptions about how an interaction should unfold} \cite{Burgoon2005}. These frameworks imply that stylistic variation in conversational agents \dk{may affect} how interactions are perceived and evaluated, shaping impressions of warmth, competence, social presence, and also eliciting linguistic alignment or accommodation \cite{Sterken2024}.

Furthermore, prior research \nuria{suggests} that the stylistic appropriateness of conversational agents is context-dependent: informal or friendly styles can increase \dk{perceived} social presence, yet may also be \dk{judged} as unsuitable in highly goal-directed tasks \cite{liebrecht2020too}. Moreover, perceived ``human-likeness'' in conversational agents appears to rely less on warmth than on clarity, consistency and linguistic precision \cite{Svenningsson2019}. While \dk{prior} research has examined dimensions such as empathy, politeness and anthropomorphism in conversational agents \cite{bickmore2005establishing}, relatively few studies have systematically isolated \emph{fixed stylistic tone} in \dk{a} controlled, goal-oriented \dk{task} where performance can be objectively measured while \dk{task} content is held constant. Moreover, it remains unclear whether linguistic accommodation emerges in short, tightly constrained interactions, or whether user characteristics --such as gender or prior experience with chatbots-- \dk{condition responses to} communication style. Recent findings further suggest that user responses to stylistic variation depend on contextual appropriatedness and the users' awareness of interacting with an artificial agent \cite{islind2023friendly}, underscoring the need for controlled experimental evidence.

In this paper, we focus on chatbots as one class of conversational agents. We investigate how a chatbot's communication style influence task performance, interaction behavior, and subjective user evaluations within a \dk{controlled} map-based 2D navigation task. To this end, we design two stylistic variants of a chatbot \naviw: a friendly and supportive style (\naviw$_f$) or a direct, task-focused style (\naviw$_d$). \dk{These two chatbot conditions} were compared with a control condition in which participants \dk{received static step-by-step instructions without a chatbot}. In addition, we investigate whether users linguistically accommodate to the chatbot's communication style and whether such effects are moderated by individual differences, including prior experience with chatbots and gender. \dk{The design allows us to examine how stylistic tone relates to subjective satisfaction, task success and linguistic behavior while keeping the navigation route and task goal constant.}

Our contributions are threefold: (1) we provide evidence on how fixed communication style relates to both relational (communication satisfaction) and instrumental (task success) in a measurable, controlled and goal-oriented task; (2) we \dk{examine whether users linguistically accommodate to the chatbot's style} in short, task-focused exchanges; (3) \dk{we study whether gender and prior chatbot experience shape responses to style}, with implications for the design of inclusive and socially attuned conversational agents.

The rest of the paper is structured as follows: we provide a summary of the most relevant related work and formulate the research questions that drive our research in Section~\ref{sec:relwork}. Our user study and its results are presented in Section~\ref{sec:userstudy} and Section~\ref{sec:results}, respectively. We provide a discussion of our findings in Section~\ref{sec:discussion}, followed by limitations and future work in Section~\ref{sec:limitations}, and conclusion in Section~\ref{sec:conclusion}.

\section{Related work} 
\label{sec:relwork}

\subsection{Communication style in conversational agents}

Research in HCI demonstrates that users respond socially to systems that employ human-like cues \cite{reeves1996media,nass2000machines}. Conversational style, \dk{understood as the patterned use of} tone, politeness, affective framing, \dk{directness and interactional support in dialogue,} influences the user's perceptions of a conversational agent's personality, appropriateness and competence \cite{Feine2019,mairesse2011controlling,pralat2024}. Friendly or warm styles tend to increase perceived social presence and relational engagement, whereas more direct or task-focused tones may enhance perceived clarity \dk{and} efficiency \dk{while reducing perceived} warmth \cite{Go2019,bhattacharjee2025,deng2025}. \dk{However,} stylistic effectiveness \dk{is not universal:} informal or friendly styles can increase social presence but may be judged as \dk{inappropriate in highly} instrumental contexts \cite{liebrecht2020too}, and perceived humanness in task-oriented chatbots \dk{may depend strongly on} linguistic precision rather than emotional expressiveness \cite{Svenningsson2019}.

\dk{This literature suggests that communication style should be treated as a design parameter whose effects depend on task demands and user expectations.} From a socio-cognitive perspective, warmth and competence are two central dimensions of social judgment \cite{Fiske2007}. In human–AI interaction, warmth has been linked to affective trust and satisfaction, while competence supports performance-related trust \cite{Kulms2018,Lee2004}. The broader quality of communication experience, including clarity, responsiveness and relational cues, offers an additional lens for understanding stylistic effects \cite{Gray2004}. 
\dk{Thus, a friendly style may improve subjective communication experience without necessarily improving objective task performance, whereas a direct style may be efficient but less relationally satisfying. This distinction is central for the present study.}

In addition, most prior work has examined adaptive, affective or multimodal agents \cite{gnewuch2018faster,Lee2011}. Less attention has been given to fixed, non-adaptive styles in controlled tasks where objective outcomes can be measured \dk{and where the informational route is held constant} \cite{qin2025,diederich2022}. This limits our understanding of \dk{whether} stylistic tone alone \dk{affects} the users' subjective evaluations of the interaction, and whether such effects extend to task performance. \dk{It also matters whether users know they are interacting with a chatbot rather than a human, because this awareness can shape expectations and evaluations} of style \cite{islind2023friendly}. 
\dk{The present study focuses on a chatbot that was explicitly presented to participants as a conversational agent and manipulates its communication style.}

\subsection{Social alignment and linguistic accommodation}

Communication Accommodation Theory (CAT) describes how individuals adjust lexical, syntactic and paralinguistic behavior to manage affiliation or distinction \cite{giles2023communication, giles2013communication}. Evidence of linguistic alignment with artificial interlocutors has been observed in text-based and voice-based interactions, including mirroring of pronouns, affective expressions and syntactic structures. However, such accommodation is often selective and feature-specific rather than reflecting global convergence \dk{of style} \cite{luger2016like}. Expectancy-driven models \dk{similarly} propose that communicative \dk{behavior is} shaped by anticipated interactional \dk{norms, with} alignment or compensation \dk{depending} on whether the agent's style confirms or violates these expectations \cite{Burgoon2005}. Congruence between agent tone and user expectations \dk{can strengthen} rapport and social presence, whereas mismatches can hinder cooperation \cite{ruijten2015lonely, Feine2019, Nass1996}.

\dk{The practical relevance of linguistic accommodation is that it helps distinguish two kinds of stylistic influence. A chatbot's style may affect how users evaluate the interaction, but it may or may not shape how users themselves communicate during the interaction. If users accommodate to an agent's style, stylistic choices can alter the interaction dynamics and not only post-task impressions. If accommodation is limited, style may operate mainly through perception, satisfaction and task engagement. Most existing evidence comes from longer, richer or multimodal interactions. It therefore remains unclear whether accommodation emerges in short, goal-oriented text exchanges where users focus primarily on solving a task.}

\subsection{Gender and relational orientation in HCI}
Sociolinguistic and psychological research \dk{has reported average gender-related} differences in communication \dk{goals, with women often described as more oriented toward} relational and affiliative \dk{cues and men as} more \dk{oriented toward} instrumental aspects such as efficiency or control \cite{tannen1990you, cross1997models}. In technology use, \dk{prior work similarly suggests that} women tend to be more sensitive to socially expressive interfaces, while men evaluate systems more through usefulness and ease of use \cite{Sobieraj2020,gefen1997gender}. Studies further show that women experience higher social presence and relational engagement \nuria{when interacting with} polite or friendly agents \cite{ruijten2015lonely}, suggesting that communication style may interact with gender to shape satisfaction and trust \cite{ruijten2015lonely}.

\dk{At the same time, gender-related differences in HCI are typically context-dependent, can be impacted by task design and prior experience \cite{nazareth2019meta}, and may not appear as statistically reliable moderation effects in a given experiment. For this reason, the present study examines gender as a potential user characteristic that may condition responses to style, but we treat gender-stratified patterns cautiously and distinguish them from confirmed interaction effects.}

\subsection{Effects of prior chatbot experience on expectations and evaluations}
Beyond immediate stylistic cues, prior experience with conversational agents may shape how users perceive and evaluate chatbot interaction. Users with extensive prior experience tend to develop more calibrated mental models of \dk{system} capabilities, which can lead to higher task efficiency but also more critical assessments of system performance and language quality \cite{luger2016like}. In contrast, less experienced users \dk{may} attribute greater intelligence or intentionality to chatbots, resulting in higher initial satisfaction but also increased risk of expectation violation when system limitations become apparent \cite{rheu2024}. These differences suggest that prior chatbot experience \dk{may condition} how communication style and system behavior are perceived.

Empirical studies further indicate that prior experience influences how users evaluate conversational style, social presence and trustworthiness. Experienced users \nuria{tend to} prioritize efficiency, clarity and goal alignment over relational or anthropomorphic cues, whereas novice users may respond more positively to friendly or supportive language that scaffolds the interaction and reduces uncertainty \cite{islind2023friendly}. Experience-related differences can \dk{also} affect users' willingness to adapt their own language, with experienced users showing less linguistic accommodation and greater reliance on concise, task-oriented commands \cite{bell2003,zellou2024}. Finally, longitudinal evidence suggests that repeated exposure and regular interactions with  conversational agents influence perceptions of trust, dialogue quality, anthropomorphism and interaction behavior \cite{Araujo2024}.

\subsection{Research Questions}\label{subsec:research_questions}

Guided by Social Response Theory, Communication Accommodation Theory and socio-cognitive models of warmth and competence, the present study examines how fixed communication style affects user performance, experience and \dk{linguistic behavior} when other interactional factors are held constant. We investigate the following research questions:

\paragraph{RQ1 - Communication style and user outcomes}
How does a chatbot’s communication style (friendly vs. direct) influence task performance and satisfaction in a goal-oriented map-based 2D navigation task? This question targets both relational and instrumental effects of stylistic tone.

\paragraph{RQ2 – Moderation by user characteristics}
Do user characteristics, such as gender and previous chatbot experience, moderate the influence of communication style on task and experience measures? Given prior work on relational versus instrumental orientations \dk{and chatbot expectations,} we examine \dk{whether} sensitivity to stylistic tone \dk{differs across user groups.}

\paragraph{RQ3 – Linguistic accommodation}
Do users linguistically adapt to the chatbot’s communication style during short, task-focused interactions? By analyzing lexical and stylistic features in user messages, we test whether \dk{style affects the users' own language or primarily their evaluations of the interaction.}

We address these three research questions by means of a between-subject user study designed to evaluate communication style as both a social cue and a functional design parameter, and to assess its \dk{relationship} with user characteristics and linguistic behavior.

\section{User study} 
\label{sec:userstudy}
\subsection{Task: Map-based 2D navigation as a testbed for communication style}

Navigation tasks offer a structured environment for examining how users integrate external guidance into goal-oriented action. Unlike open-ended conversational settings, map-based 2D navigation constrains both the problem and the interaction space, thereby allowing stylistic tone to be manipulated independently of content and task structure. Prior work suggests that navigation performance is sensitive to cue structure, time pressure, and the degree of required initiative, \emph{e.g.,} distinguishing stepwise instruction-following from dialogue-based exploration \cite{dahmani2023sex,Bruny2017}.

\begin{figure}[ht]
    \includegraphics[width=\linewidth]{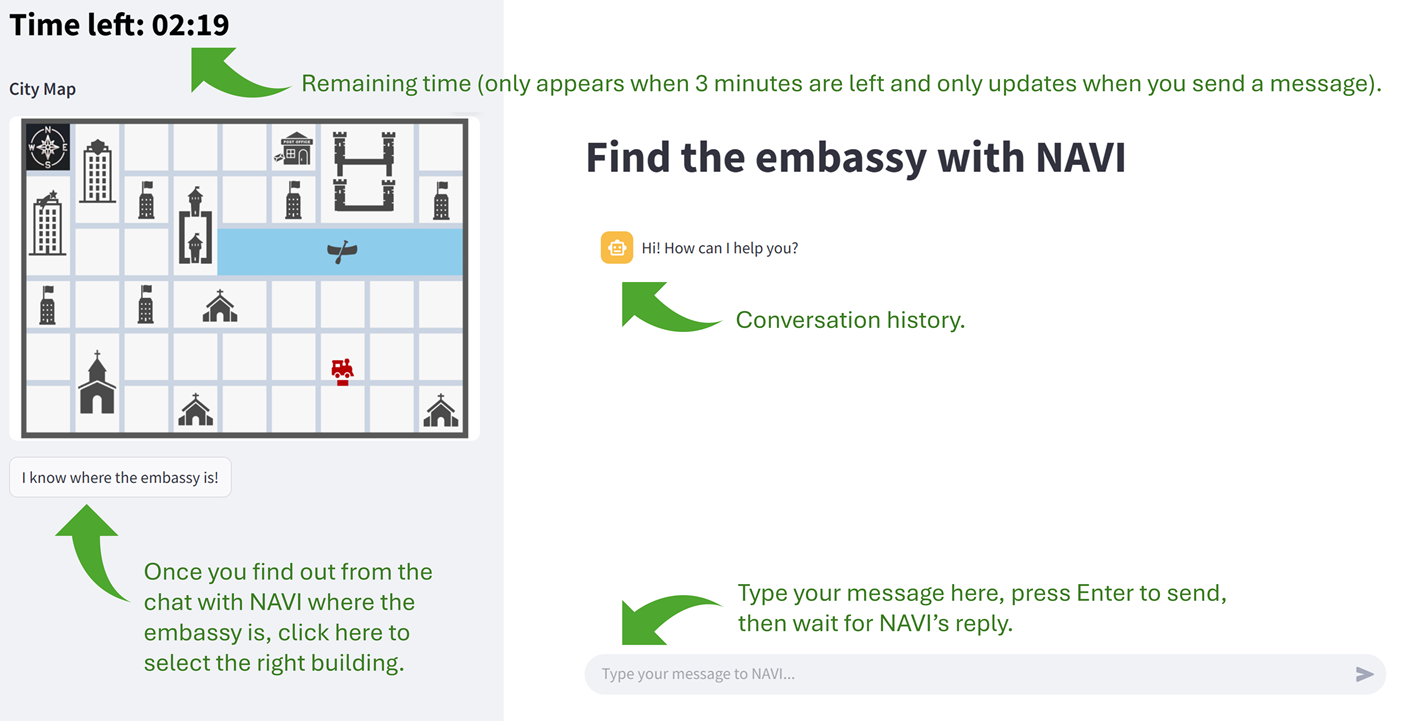}
    \caption{Explanations of the user interface used by the participants to interact with \naviw.}
    \label{fig:interface-navi}
\end{figure}

A controlled navigation task therefore offers a tractable experimental paradigm to study communication style as the primary independent variable, while holding constant task complexity, interaction modality, and informational content. To this end, we designed a map-based 2D navigation task in which participants were required to locate a predefined target destination (the embassy) using concise, content-equivalent textual guidance (See Fig. \ref{fig:interface-navi}).

In addition, we included a control condition in which participants received only step-by-step navigation instructions without interacting with a chatbot. This control condition served two purposes: it provided a baseline level of task performance in the absence of conversational interaction; and it allowed us to confirm that the navigation task itself did not systematically provide an advantage to particular user profiles. 
Note that the control condition was not intended to serve as a stylistic baseline, but rather as a validation of the task environment. Thus, comparisons between the two chatbot conditions allow for a focused assessment of the impact of communication style while holding all other task characteristics constant.

\begin{figure}[ht]
\centering
\includegraphics[width=\textwidth]{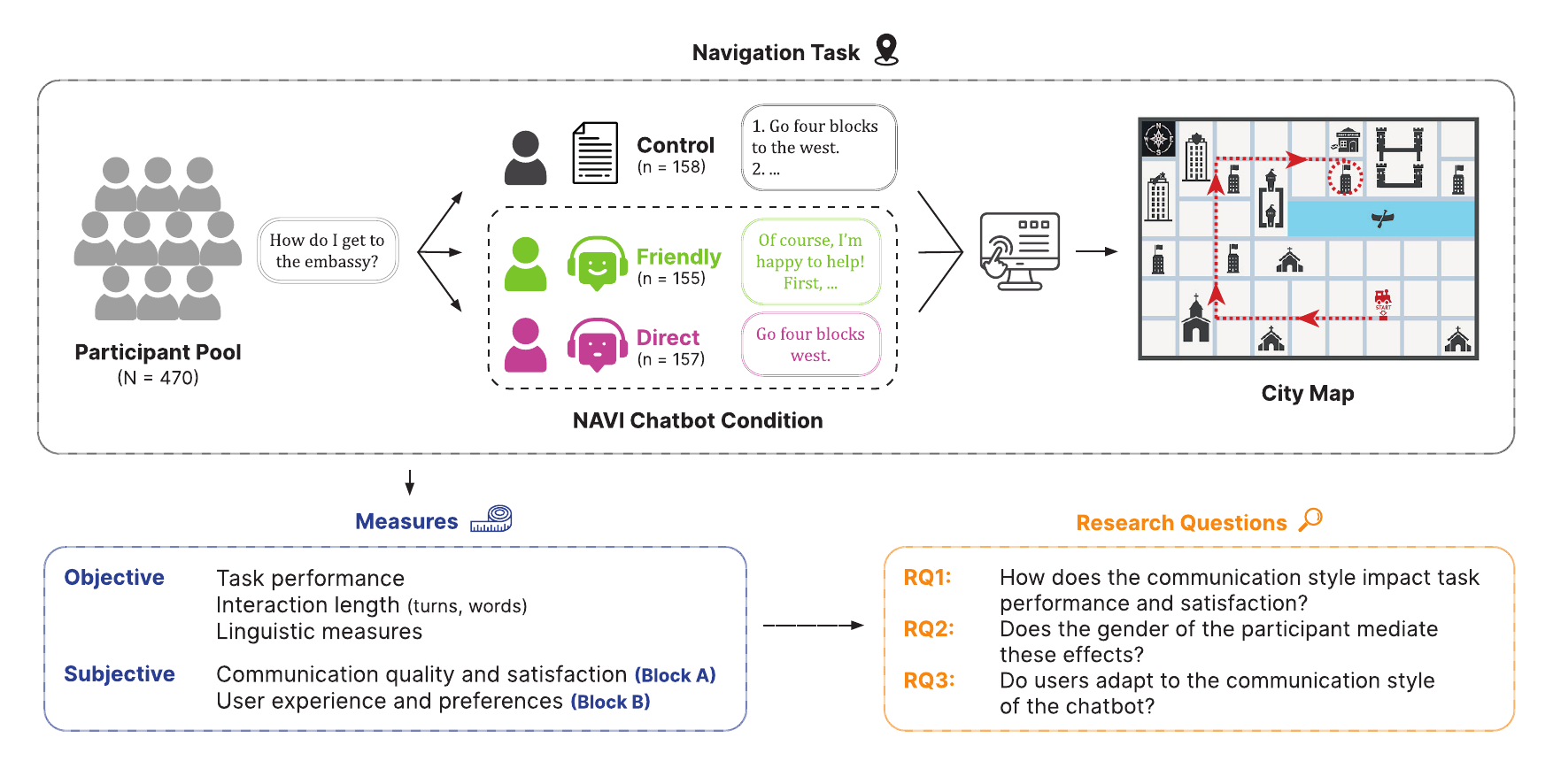}
  \caption{Overview of the study design. Participants completed a fictional map-based 2D navigation task to locate an embassy on a map (top). They followed static instructions (\textbf{Control}) or they interacted with one of the two versions of the chatbot \naviw, differing only in the communication style: \textbf{Friendly} (supportive, warm) or \textbf{Direct} (concise, task-focused). Both objective and subjective measures (bottom-left) were collected to address three research questions (bottom-right).}
  \label{fig:overview}
  \end{figure}

\subsection{Design}

We used a between-subjects experimental design, depicted in Figure \ref{fig:overview}. The study consisted of two parts: (1) a navigation task and (2) a post-task survey that collected subjective feedback regarding communication quality, satisfaction, user experience and preferences for chatbot versus human assistance. Measures are described in detail in \ref{sec:measures}, and full item wordings are provided in \ref{sec:appendix.questions}. 

\begin{figure}[ht]
\centering
    \includegraphics[width=\linewidth]{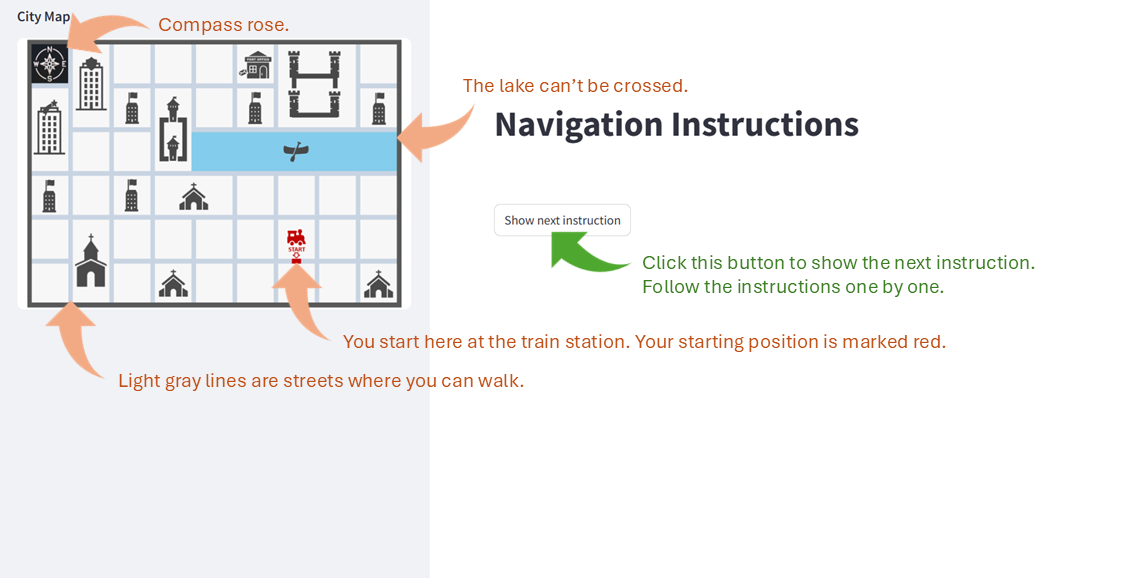}
    \caption{Explanations of the user interface presented to the participants in the control condition, including the map legend.}
    \label{fig:interface-static}
\end{figure}

\subsubsection{Map-based 2D navigation task}

Participants were instructed to imagine arriving at the train station of a capital city where they had never been before. They realized that they did not have their phone and documents with them, so they needed to urgently reach their country’s embassy. Their task was to figure out where the embassy was located by navigating from the train station to the embassy on a 2D map.

The interface displayed a static city map with streets, buildings and a large lake, together with a legend and compass rose indicating cardinal directions. Streets were represented by light grey lines that could be used for walking, the lake could not be crossed and the starting position at the train station was marked with a red square. A detailed explanation of the interface provided participants with all necessary information about the map layout, clickable elements and interaction flow, as can be seen in Figure \ref{fig:interface-static}. 

\paragraph{Control condition (no chatbot)}

In the control condition, participants received step-by-step written instructions guiding them from the train station to the embassy. The full set of instructions is provided in \ref{sec:appendix.instructions}. The interface showed a static 2D city map on the left and a panel with navigation instructions on the right, as depicted in Figure \ref{fig:interface-static}. Participants clicked a button to reveal the next instruction in sequence and were asked to follow these instructions on the map until they believed they had identified the embassy. The layout and visual elements were matched as closely as possible to the chatbot interface to ensure that the main difference between conditions was the presence of a conversational agent rather than differences in visual design or information structure.

\paragraph{Chatbot condition (\naviw)}

In the chatbot condition, participants were introduced to \naviw, a ChatGPT-based conversational agent with local knowledge that could help them find the embassy. \navi was embedded in a custom web interface that combined a chat window on the right with the same static 2D city map on the left, as illustrated in Figure \ref{fig:interface-navi}. \nuria{As shown in the Figure, \naviw's icon was a robot to indicate that participants were interacting with a chatbot}. Participants were told that they could ask \navi for directions and clarifications in order to locate the embassy. 

Upon entering the chatbot branch of the study, participants were randomly assigned to one of two \nuria{communication-style} variants of \naviw:
\begin{itemize}
    \item Friendly \navi (\naviw$_f$), which was configured to use a warm, polite and supportive \nuria{communication style}. It offered \dk{encouragement,} reiterated instructions when participants appeared \dk{confused, and would occasionally include} brief \nuria{contextual remarks (\emph{e.g.,} about nearby landmarks) alongside the same route guidance. These remarks were intended to foster rapport and social presence rather than provide additional task-relevant navigation information.}
    \item Direct \navi (\naviw$_d$), which provided brief, task-focused responses that were efficient and to the point. This version avoided social niceties, did not offer \dk{encouragement} and \dk{minimized contextual or relational commentary.} The goal was to maximize clarity and efficiency, \nuria{emphasizing instrumental communication over social engagement while conveying the same core navigation guidance.}
\end{itemize}

\nuria{The friendly communication operationalized friendliness through a broader set of conversational behaviors commonly associated with warm, rapport-oriented interactions, including a warmer tone and occasional local tips and contextual remarks while keeping the core navigation content consistent across conditions. Therefore, the manipulation reflects differences in overall communication style rather than differences in linguistic form alone.}

Participants were not informed which version of \navi they were interacting with, and both versions followed the same underlying navigation route from the train station to the embassy. In all chatbot conditions, \naviw’s behavior was controlled by style-specific system prompts, which specified communicative goals, politeness level, affective tone and constraints on content. The exact wording of the system prompts for \naviw$_f$ and \naviw$_d$ is provided in \ref{appendix:system_prompts}. To \dk{keep} the manipulation \dk{focused on} communication \dk{style,} we constrained the chatbot to reveal a fixed route step by step. The sequence of directions and landmarks was held constant across both styles, and \navi was instructed to avoid giving away the answer directly, for example, by naming the embassy building number.

Because the underlying model is generative, individual responses could still exhibit some stochastic variability. To verify that the stylistic manipulation was stable and implemented as intended, we later profiled \naviw’s utterances using the Linguistic Inquiry and Word Count (LIWC) dictionary and complementary embedding-based measures (see Section \ref{sec:measures}). These analyses confirmed that the friendly and direct versions displayed distinct and consistent patterns in affective and relational categories, while within-style variation remained low relative to between-style differences.

Participants interacted with \navi via the chat interface to obtain directions from the starting point (the train station) to the target (embassy). The interface displayed a timer that became visible only when the remaining time dropped below three minutes, and this timer updated whenever participants sent a message. In the instructions, participants were told that, on average, six to eight conversational turns were usually needed to locate the embassy. To discourage attempts to solve the task without meaningful interaction, a minimum of three user–\navi turns were enforced before participants could submit their final answer with the location of the embassy.

In both the control and chatbot conditions, participants indicated that they had reached a decision by clicking a button located below the map on the left-hand side of the interface with the label \emph{“I know where the embassy is”}. Once they clicked this button, numbers from 1 to 15 appeared over all buildings on the map. Participants were then asked to select the number corresponding to the building they believed to be the embassy. Feedback about correctness was not provided at any point to avoid learning or feedback effects.

The navigation task had a maximum duration of eight minutes. If a participant had not submitted an answer by the end of this period, the task was terminated and the participant was automatically advanced to the post-task survey. The eight-minute limit was chosen based on pilot data, which indicated that this duration was sufficient for participants to complete the task under normal circumstances while still imposing a realistic sense of time pressure.

\subsubsection{Post-task survey} After completing the navigation task and selecting the embassy building, participants proceeded to a post-task survey. The survey captured subjective evaluations of the interaction and provided additional context for interpreting objective task measures.

In the chatbot conditions, participants completed a questionnaire with randomized block order and embedded attention checks. The survey contained questions about perceived communication quality and satisfaction, perceived usefulness of \navi for accomplishing the task, and preferences for interacting with a chatbot versus a human in similar situations. It also captured chatbot use frequency to contextualize individual differences in user experience. The complete set of survey items, including the wording and response scales, is provided in \ref{sec:appendix.questions}, and all derived measures are described in Section \ref{sec:measures}.

In the control condition, participants completed a shorter post-task survey that did not include communication quality items, since they had not interacted with a conversational agent. Instead, they evaluated their satisfaction with the written instructions, their preference for receiving directions from a human versus written instructions, their preference for using a chatbot instead of static instructions and their chatbot use frequency. The corresponding items for the control condition are listed in \ref{sec:appendix.questions} (control survey section), again with measures described in Section \ref{sec:measures}.

\subsection{Implementation} The study was implemented using Streamlit, an open-source Python framework for interactive web applications. Streamlit allowed us to integrate instructions, the map interface, the chat window and the post-task survey within a single application, which reduced context switching for participants and simplified logging of interaction data. The application was hosted on Microsoft Azure, which provided scalable and reliable access for Prolific participants. Azure storage was used to securely log each participant’s progression through the study, including navigation behavior, chat transcripts and survey responses. The chatbot was powered by the gpt-4.1-2025-04-14 model via the OpenAI API, which served as the backbone for generating \naviw's responses. The study was integrated with Prolific’s recruitment and screening tools, enabling automatic enforcement of inclusion criteria and controlled access to the study link.

Participants were randomly assigned by the web application to one of the three experimental conditions (control, friendly \naviw, direct \naviw). Within the chatbot branch, assignment to friendly versus direct \navi was based on simple random allocation with equal target probability. To maintain approximately equal numbers of women and men in each condition, we additionally monitored recruitment through Prolific’s prescreening and quota tools and adjusted ongoing recruitment so that gender distributions remained balanced across conditions.

\subsection{Pilot studies} \label{subsec:pilot}
To refine the study design and ensure a smooth user experience, we conducted three iterative pilot studies on Prolific, each with 10 participants. After each pilot, we examined navigation performance, chat transcripts and survey responses, and adjusted the interface and protocol accordingly. Several key design choices were informed by these pilots:
\begin{itemize}
    \item We introduced a minimum number of required conversational turns, because early pilots showed that some participants tried to guess the embassy location without interacting meaningfully with \naviw.
    \item We refined the time estimates and compensation to align with the actual time needed to complete the task and survey, and we updated the information provided to participants so that expectations about study duration were realistic.
    \item We clarified and streamlined the instructions, including the narrative scenario, explanation of the interface and examples of how to ask \navi for help.
    \item We adjusted the color scheme and contrast of the map to improve readability and reduce visual ambiguity. 
\end{itemize}
    
While the map layout (landmark positions) and the style-specific system prompts were initially developed through exploratory prototyping, both were fixed prior to the pilot phase and remained unchanged across all pilot and main study sessions. The final version of the instructions shown to participants is included in \ref{sec:appendix.instructions}.

\subsection{Sample size and power considerations} \label{subsec:sample_size}
Before data collection, we conducted \emph{a priori} power considerations to determine the sample size. Prior work on stylistic manipulations in conversational agents typically reports effects in the small-to-medium range on relational outcomes such as trust and satisfaction, with Cohen’s d values often between $0.25$ and $0.40$. Assuming a two-sided $\alpha$ of $.05$ and desired power of $.80$, a medium effect (d $\approx 0.50$) can be detected with about 64 participants per group, whereas detecting a lower-medium effect (d $\approx 0.30$) requires larger groups of around n = 170–180 participants per condition. For correlations between linguistic indicators and subjective evaluations, similar calculations show that medium correlations (r $\approx 0.30$) require approximately n = 80–90 participants, while smaller correlations (r $\approx 0.20$) require about n = 190–200 participants. Anticipating effects in the small-to-medium range and aiming to support both between-condition comparisons and exploratory analyses of linguistic alignment and gender moderation, we targeted a total sample of approximately $450$ to $500$ participants distributed across the three conditions. The final sample of $470$ participants (see \ref{subsec:participants}) thus provides adequate sensitivity for detecting stylistic effects of theoretical interest and sufficient stability for the text-based analyses.

\subsection{Ethical approval} \label{subsec:ethical}
The full study protocol, including recruitment, consent materials, data handling and debriefing procedures, was reviewed and approved by the ethics committee of the Faculty of Education, University of South Bohemia (approval reference number: EK049/2025, dated April 28, 2025). The committee concluded that the project complied with institutional regulations and international standards for research with human participants. All participants provided informed consent prior to beginning the study and were compensated in line with Prolific guidelines and the estimated study duration.

\begin{figure}[htbp]
    \centering
    \begin{subfigure}[t]{0.58\textwidth}
        \includegraphics[width=\linewidth]{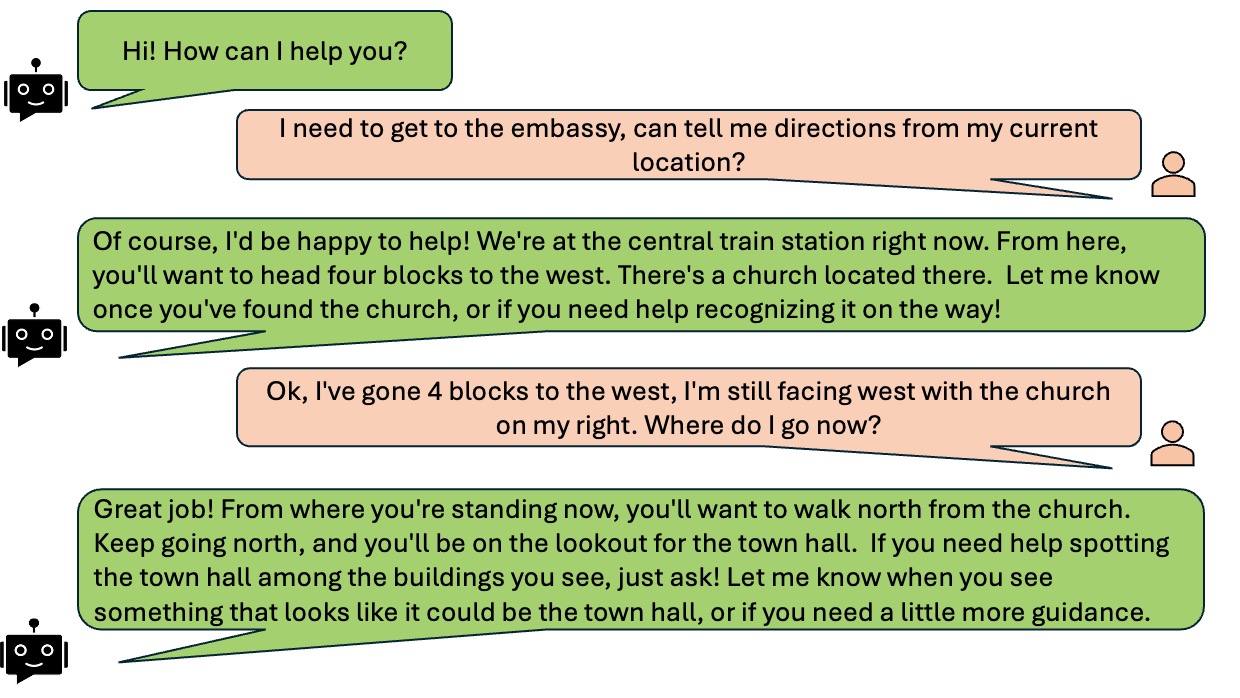}
        \caption{Friendly \navi}
        \label{fig:sampleConvos.friendly}
    \end{subfigure}%
    \hfill
    \begin{subfigure}[t]{0.38\textwidth}
        \includegraphics[width=\linewidth]{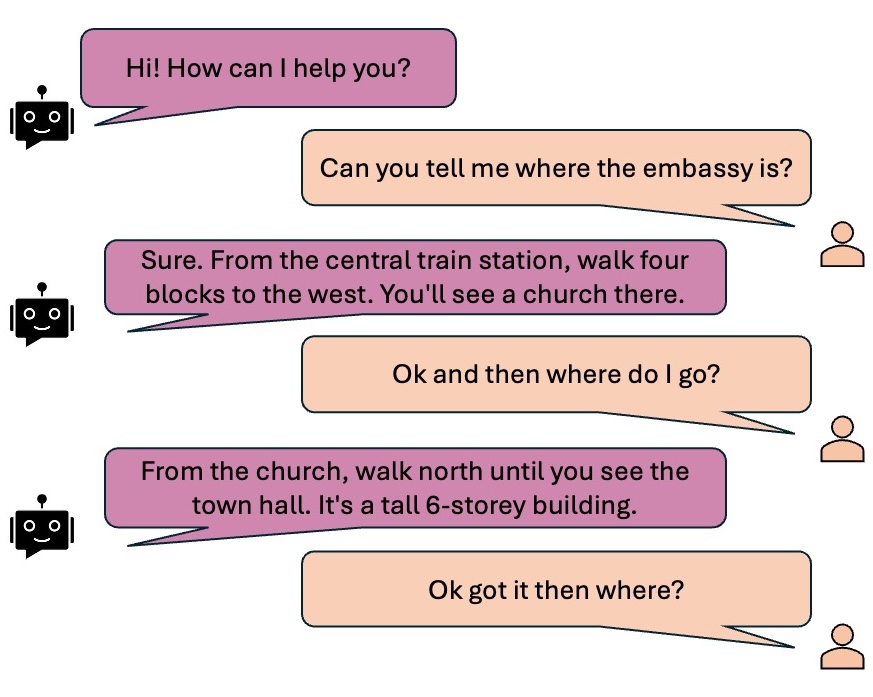}
        \caption{Direct \navi}
        \label{fig:sampleConvos.direct}
    \end{subfigure}%
    \caption{Excerpts of exemplary conversations with the friendly (left) and direct (right) versions of \naviw.}
    \label{fig:sampleConvos}
\end{figure}

\subsection{Measures}
\label{sec:measures}

We collected objective indicators of task performance and interaction behavior, self-reported evaluations of the interaction, and linguistic measures capturing structural, stylistic and affective aspects of the interaction. This combination of measures was designed to address our three research questions by jointly characterizing instrumental outcomes (task success and persistence), relational outcomes (satisfaction) and linguistic accommodation.

\subsubsection{Objective Measures}
We collected the following objective measures: 

\textbf{1. Task performance}, defined as a binary indicator of whether participants correctly identified the embassy building at the end of the navigation task. This measure directly captures the primary instrumental outcome of the guidance and provides a simple, interpretable criterion for comparing conditions.

\textbf{2. Interaction behavior}, measured by the number of conversational turns between the participant and \navi, the number of explicit help requests and the total time spent on the navigation task. Because reaction-time data collected in uncontrolled online environments are often noisy, we treated time as a secondary contextual measure rather than a primary efficiency metric. The main behavioral indicators were therefore the extent of conversational engagement (number of turns) and reliance on the guidance (number of help requests).

\textbf{3. Linguistic measures}, to assess linguistic accommodation, consisting of three variables that measure structural (length), stylistic (embedding-based) and affective (valence) dimensions of linguistic alignment. 
\begin{enumerate}
    \item \emph{Message length alignment}, measured by the difference in average message length between the messages produced by \navi and those generated by the participant. Note that convergence in word count is a standard structural marker of linguistic alignment.
    \item \emph{Stylistic similarity in the embedding space}, quantified using cosine distances between the participant's and \naviw's utterances in sentence-level style embeddings developed by Wegmann \emph{et al.} \cite{Wegmann2022}, which capture stylistic features independently of topical content.
    \item \emph{Overlap in positively valenced vocabulary}. As \naviw$_f$ was designed to use more positively valenced wording than \naviw$_d$, we computed the overlap in affectively positive words using a RoBERTa-based valence estimation model \cite{mendes2023}. Words with valence > 0.6 were classified as positively valenced. 
\end{enumerate}
    
Furthermore, a descriptive lexical analysis of utterances using Linguistic Inquiry and Word Count dictionary (LIWC) \cite{pennebaker1999linguistic} was conducted to document the linguistic patterns of the language used by both \navi and the participants. LIWC assigns words to psychologically meaningful categories (\emph{e.g.,} pronouns, affect, cognition). Importantly, LIWC was not used as an inferential test of accommodation but as a transparent, replicable description of the linguistic material for future comparisons using dictionary-based analysis tools.

\subsubsection{Subjective Measures} 
Subjective measures were collected by means of questionnaires that were delivered after the task was completed and targeted relational and instrumental aspects of the interaction. The questionnaires were structured in two Blocks: 

\paragraph{1. Block A}
In the chatbot conditions, subjective \nuria{communication} satisfaction \nuria{was assessed} by means of the Interpersonal Communication Satisfaction Inventory (ICSI) \cite{hecht1978conceptualization}, \nuria{a validated} 19-item \nuria{questionnaire rated on a} seven-point Likert scale. It was created through systematic item generation, empirical testing, and psychometric validation. \dk{In its original validation, the inventory showed} high \dk{split-half} reliability \dk{with Spearman–Brown correction (coefficients of approximately 0.90–0.97 across conversation types and relationship levels)} and good convergent validity \dk{against faces-scale satisfaction measures} (0.64–0.87), supporting its use as a largely unidimensional measure of global communication satisfaction. For our study, the inventory was adapted by replacing \nuria{references to the} “other person” with the chatbot identifier \naviw. Sample items include: \emph{“\navi let me know I was communicating effectively”}, \emph{“The conversation flowed smoothly”}, and \emph{“I was satisfied with the conversation”}. \adi{The updated instrument demonstrated high internal consistency in our sample (Chronbach's $\alpha$=0.71), indicating that the items functioned coherently as a measure of communication satisfaction in this context. } The \nuria{complete questionnaire} can be found in ~\ref{sec:appendix.questions}. The \nuria{inventory} covers \nuria{multiple} aspects \nuria{conversational experience, including} emotional engagement, \nuria{conversational flow}, responsiveness, and mutual understanding, \dk{and contains several} reverse-coded \dk{items.} The adapted ICSI \dk{served as a theoretically grounded} measure of subjective interaction quality to complement task performance outcomes when comparing \naviw's friendly versus direct chatbot communication styles. \nuria{To preserve the psychometric structure of the validated instrument, we administered the complete inventory rather than selecting individual items.} ICSI scores were calculated according to the \dk{original} scoring procedure \cite{hecht1978conceptualization}. \nuria{Negatively worded items were reverse-coded and responses were summed to produce a total score ranging from 19 to 133, with higher scores indicating greater communication satisfaction.} Two attention \dk{checks items were included in the survey as a whole (one per block), separately from the ICSI inventory itself.}

Note that this block was omitted for participants in the control condition, as they did not interact with a chatbot and, therefore, had no communicative elements to evaluate.

\paragraph{2. Block B} In addition, participants were asked to provide feedback about their user experience and preferences by means of four 7-point Likert questions. 

In the chatbot conditions, the questions were: (1) \emph{Perceived naturalness}, assessed by how human-like and fluent the conversation went (\textit{“The conversation felt natural and human-like".}); (2) \emph{Task helpfulness}, measured by how effectively \navi supported the user in completing the task (\textit{“\navi helped me solve the task effectively".}); (3) \emph{Human preference}, which was captured by the users' subjective evaluation and overall preference for interacting with a human instead of with a chatbot (\textit{“I would have preferred to ask a human".}); and (4) \emph{Chatbot familiarity}, which measured the participants' prior experience and familiarity with chatbots, providing context for interpreting their responses (\textit{“How frequently do you use chatbots?”}).  

In the control condition, the four 7-point Likert questions were: (1) \emph{Satisfaction} with the instructions (\textit{“I am satisfied with how the instructions conveyed the information on how to get to the embassy"}); (2) \emph{Human preference}, which was captured by the participants' overall preference for interacting with a human instead of receiving the instructions in a step-by-step written form (\textit{“I would have preferred receiving the instructions to get to the embassy from a human rather than in written form".}); (3) \emph{Chatbot preference}, which was captured by the participants' overall preference for interacting with a chatbot instead of receiving the instructions (\textit{“I would have preferred to ask a chatbot for help rather than follow written instructions".}); and (4) \emph{Chatbot familiarity} which measured the participants' prior experience and familiarity with chatbots, providing context for interpreting their responses (\textit{“How frequently do you use chatbots?”}).

\paragraph{3. Attention checks} Furthermore, there were two attention checks, one per block, presented in the middle of the block, where participants were asked to select a specific option on the Likert scale (\emph{e.g.}, \textit{``For this question, ignore all the instructions and select `Disagree'.''}). Participants had to pass both attention checks to be included in the study.

\ref{sec:appendix.questions} contains all the questions that participants responded to in this part of the study.

\subsection{Participants}\label{subsec:participants}

Participants were recruited via Prolific and were required to be based in the United States. We obtained a gender-balanced sample and restricted participation to individuals with an approval rate between 85–100\% on Prolific. To minimize potential confounds, we further limited the sample to neurotypical individuals whose first language was English, excluding those reporting neurodiversity, dyslexia, or vision impairments.

After applying exclusion criteria for failed attention checks and statistical outliers, a total of 470 valid participants took part in the study. The control condition included 158 participants (79 female, 79 male). The chatbot condition included 312 participants, divided between two versions of \naviw: the direct version (N = 157, 78 female, 79 male) and the friendly version (N = 155, 78 female, 77 male). The resulting sample size aligns with our a priori target (see Section \ref{subsec:sample_size}) and provides reasonable sensitivity for detecting small-to-medium stylistic effects as well as stable estimation in the linguistic analyses.
Gender representation was approximately balanced in each condition. 

The target compensation was set to USD~12 per hour, following Prolific's fair-pay guidelines. From the pilot studies, we estimated the study to take 12 minutes in the chatbot condition and 5 minutes in the control condition. On average, participants completed the study in 10 minutes in the chatbot condition and in 6 minutes in the control condition. All participants provided informed consent before starting the study.

\section{Results}
\label{sec:results}

In this section, we present the results of analyzing the data collected in the user study to address the research questions formulated in Section~\ref{subsec:research_questions} by means of the collected measures. For all statistical analyses, the significance threshold was set at $\alpha = 0.05$. Prior to conducting comparisons, the normality of the distributions was assessed using the Kolmogorov–Smirnov test, and the homogeneity of variances was evaluated with Levene's test. When both assumptions were satisfied, independent samples t-tests were applied; otherwise, non-parametric Kruskal--Wallis and Mann–Whitney tests were used. The results reported below provide the empirical basis for our discussion of the broader implications, which we develop in Section~\ref{sec:discussion}.

\subsection{RQ1: Communication style and user outcomes}

Table \ref{tab:icsi_style} depicts the linear regression conducted to examine whether communication style predicted the ICSI scores \emph{i.e.,} the participants' satisfaction with their conversation. The model was statistically significant, F(1, 310) = 10.28, p < 0.01, though it explained a modest proportion of variance (adjusted $R^2$ = 0.029, \adi{95\%BCa CI [0.0016,0.0756]}).

\begin{table}[ht]
\centering
\begin{tabular}{lccccc}
\toprule
\textbf{Predictor} & \textbf{Estimate} & \textbf{SE} & \textbf{t} & \textbf{p} & \textbf{\adi{95\% CI}} \\
\midrule
Intercept         & 95.860 & 1.321 & 72.56 & < 2e-16 & \adi{[93.26, 98.46]} \\
Style (Friendly)  & 6.011  & 1.874 & 3.21  & 0.0015 & \adi{[2.32,9.70]} \\
\midrule
\multicolumn{6}{l}{Residual SD = 16.55; df = 310} \\
\multicolumn{6}{l}{Multiple $R^2$ = 0.032; Adjusted $R^2$ = 0.029; $F$(1, 310) = 10.28, $p$ < 0.01} \\
\bottomrule
\end{tabular}
\caption{Linear regression fit to the ICSI scores provided by participants to assess whether communication style predicted their satisfaction with the conversation.}
\label{tab:icsi_style}
\end{table}

Participants interacting with \naviw$_f$ \nuria{reported significantly higher communication satisfaction} than those interacting with \naviw$_d$ (b = 6.01, SE = 1.87, t = 3.21, p < 0.01). \nuria{Based on the fitted model,} the predicted ICSI score increased from 95.86 for \naviw$_d$ to 101.87 for \naviw$_f$, \nuria{corresponding to an improvement of approximately six points on the 19-133 ICSI scale (approximately Cohen's \adi{$d = 0.36$ [0.14,0.59]}, a small-to-moderate effect). Although the effect accounted for a modest proportion of the variance in communication satisfaction, it indicates that a friendlier communication style consistently improved the participants' subjective evaluation of the interaction.} \adi{By contrast, communication style had no statistically significant effect on the remaining subjective measures collected in Block B of the survey. These results are reported in \ref{sec:appendix.impactCommunicationStyle}.}

\begin{table}[ht]
\centering
\begin{tabular}{lccccc}
\toprule
\textbf{Predictor} & \textbf{Estimate} & \textbf{SE} & \textbf{z} & \textbf{p} & \textbf{\adi{95\% CI}} \\
\midrule
Intercept         & 0.480 & 0.164 & 2.93 & 0.0034 & \adi{[0.162,0.807]} \\
Style (Friendly)  & 0.644 & 0.249 & 2.59 & 0.0096 & \adi{[0.161,1.14]} \\
\midrule
\multicolumn{6}{l}{\textit{Model fit:} Null deviance = 388.34 (df = 311)} \\
\multicolumn{6}{l}{Residual deviance = 381.50 (df = 310); AIC = 385.5} \\
\bottomrule
\end{tabular}
\caption{Logistic regression predicting task success from communication style. The binomial model estimates the log probabilities.}
\label{tab:taskSuccess_style}
\end{table}

A binary logistic regression was performed to examine whether communication style predicted task success. Style was coded with the direct condition as the reference category. The reduced residual deviance, as indicated in Table \ref{tab:taskSuccess_style}\dk{, suggests} that including style \dk{improved model fit relative to the null model.}

The effect of communication style was statistically significant. Participants exposed to \naviw$_f$ were more likely to succeed than those interacting with \naviw$_d$, b = 0.64, SE = 0.25, z = 2.59, p < 0.01 , \adi{OR = 1.90, 95\% CI [1.17, 3.12]}. Applying the inverse-logit transformation to the model coefficients indicated that the predicted probability of task success increased from approximately $0.62$ [0.54,0.69] in the direct condition to $0.76$ [0.68,0.82] in the friendly style condition, \adi{corresponding to an absolute difference of 0.14 [0.04,0.24] in task success across styles.} 

\definecolor{mygray}{HTML}{E5E5E5}
\begin{tcolorbox}[breakable,colback=mygray!30]
Overall, the two models indicate that participants were more satisfied when engaging with \naviw$_f$, and they also exhibited higher task success rates in the friendly condition.
\end{tcolorbox}

\subsection{RQ2: Moderation by user characteristics}

In this section, we study to which degree gender and previous experience with chatbots \dk{moderated} differences in task performance and subjective satisfaction. \dk{Given that ``gender'' refers to the participants' binary self-reported sex item collected via Prolific (see Section~\ref{subsec:participants}), we interpret the results as gender-related patterns rather than as evidence about gender identity.} To assess whether gender differences could be attributed to baseline variation in navigation ability --as suggested in prior literature \cite{lawton1994gender,nazareth2019meta}-- we conducted a control study in which participants did not interact with a chatbot but instead received step-by-step navigation instructions.

\subsubsection{Control study: baseline performance}

No significant differences in task success were observed between male and female participants ($\chi^2$(1) = 0.442, p > 0.1, Cramer's V = 0.053, \adi{95\% CI [0.00,0.21]\footnote{\adi{Percentile bootstrap with 2000 resamples}}}) in the control condition. However, gender did influence attitudes toward chatbot preference: male participants expressed a significantly stronger preference for interacting with chatbots than female participants (Mann--Whitney, W = 2548.5, p < 0.05; \adi{; rank-biserial r = 0.183, 95\% CI [0.005, 0.351])}).

The absence of significant gender differences in task success under the control condition supports the \dk{task-specific} validity of the \nuria{experimental} design. \dk{Specifically, for this 2D map navigation task, the baseline instructions} did not \dk{appear to favor either gender.} Hence, any gender-related patterns observed in the chatbot conditions \dk{are less likely reflect pre-existing differences in baseline navigation performance. Nevertheless, these findings should be interpreted alongside the full interaction models, which account for the combined effects of gender and experimental condition}.

\subsubsection{Impact of gender and prior chatbot use frequency on task performance}

To assess whether the effect of communication style on task success differed by participant gender, we first fit a logistic regression model including the additive effects of communication style and gender. In this model, the friendly conversational style significantly increased the likelihood of task success (b = 0.64, SE = 0.25, z = 2.59, p < 0.01, \adi{OR = 1.90, 95\% CI [1.17, 3.12]}), whereas gender was not a significant predictor (p > 0.05). Since prior work indicates that men and women may respond differently to conversational tone, and because descriptive patterns hinted at potential differences in success rates across style × gender combinations, we next \nuria{extended the model by} including the interaction between \nuria{communication} style and gender, together with prior chatbot use frequency as a covariate. The parameters of this logistic regression model are summarized in Table \ref{tab:taskSuccess_sexAndFrequency}. Communication style was coded with the direct condition as the reference category and Gender was coded with Female as the reference category.

\begin{table}[ht]
\centering
\begin{tabular}{lccccc}
\toprule
\textbf{Predictor} & \textbf{Estimate} & \textbf{SE} & \textbf{z} & \textbf{p} & \textbf{\adi{95\% CI}} \\
\midrule
Intercept                         & 0.973  & 0.398 & 2.44 & 0.015 & \adi{[0.20,1.77]} \\
Style (Friendly)                  & 1.050  & 0.363 & 2.89 & 0.004 & \adi{[0.35,1.78]}\\
Gender (Male)                        & 0.427  & 0.335 & 1.28 & 0.202 & \adi{[-0.23,1.09]}\\
Chatbot use frequency                 & -0.166 & 0.081 & -2.06 & 0.040 & \adi{[-0.33,-0.01]}\\
Style (Fr.) $\times$ Gender (M) & -0.747 & 0.506 & -1.48 & 0.140 & \adi{[-1.75,0.24]}\\
\midrule
\multicolumn{6}{l}{Null deviance = 388.34 (df = 311); Residual deviance = 375.47 (df = 307)} \\
\multicolumn{6}{l}{AIC = 385.47; $N$ = 312} \\
\bottomrule
\end{tabular}
\caption{Logistic regression predicting task success from communication style, gender, their interaction, and prior chatbot-use frequency.}
\label{tab:taskSuccess_sexAndFrequency}
\end{table}

The interaction model provided a better fit than the null model (Null deviance = 388.34; Residual deviance = 375.47; AIC = 385.47). As seen earlier, participants interacting with \naviw$_f$ were significantly more likely to succeed than those interacting with \naviw$_d$ (b = 1.05, SE = 0.36, z = 2.89, p < 0.01, \adi{OR = 2.86, 95\% CI [1.42,5.93]}). Gender \nuria{was not a significant predictor of} task success \adi{(p > 0.05)}. 
Importantly, the communication style × gender interaction was not statistically significant (b = -0.75, SE = 0.51, z = -1.48, p > 0.1, \adi{OR=0.47, 95\% CI [0.17,1.27]}), indicating that the effect of communication style did not differ reliably between women and men at the model level.

\nuria{Although the interaction did not provide evidence of moderation, we conducted exploratory simple-effects contrasts using estimated marginal means to characterize the pattern of estimated effects within each gender. Among women, \naviw$_f$ was associated with significantly higher task success than \naviw$_d$ (Odds Ratio = 0.35 for \naviw$_d$ vs \naviw$_f$; $z=-2.89$; $p < 0.01$), corresponding to approximately 2.9 times greater odds of success with \naviw$_f$. Among men, the corresponding contrast was not statistically significant (Odds Ratio = 0.74 for \naviw$_d$ vs \naviw$_f$; $z=-0.86$; $p > 0.1$). 
Because the overall interaction was not statistically significant, these subgroup contrasts should be interpreted as descriptive and hypothesis-generating rather than as evidence of a confirmed gender-specific effect. }

Prior chatbot use frequency (provided by participants as part of the post-task survey on a 7-point Likert scale) also had a significant impact on task performance. Interestingly, higher chatbot use frequency was associated with a reduced likelihood of success (b = -0.17, SE = 0.08, z = -2.06, p < 0.05; \adi{standardized OR = 0.77, 95\% CI [0.60, 0.99])}). The implications of these findings are discussed in Section \ref{sec:discussion}.

\subsubsection{Impact of gender on chatbot satisfaction}

To investigate which factors shaped the participants’ subjective satisfaction with the interaction, we fit a linear regression model predicting the ICSI scores from communication style, the participant's gender, their interaction, self-reported chatbot use frequency, and task success. The model showed a clear effect of communication style: participants rated the conversation as significantly more satisfying when interacting with with \naviw$_f$ than with \naviw$_d$ (b = 6.81, SE = 2.61, t = 2.61, p < 0.01), corresponding to an increase of nearly 7 points on the ICSI (\adi{adjusted Cohen's $d$ = 0.34, 95\% CI [0.11, 0.56], averaged across gender given the non-significant style $\times$ gender interaction reported below)}.

\begin{table}[ht]
\centering
\begin{tabular}{lccccc}
\toprule
\textbf{Predictor} & \textbf{Estimate} & \textbf{SE} & \textbf{t} & \textbf{p} & \textbf{\adi{95\% CI}}\\
\midrule
Intercept                          & 85.13 & 3.25 & 26.23 & <0.001 & \adi{[78.7,91.5]} \\
Style (Friendly)                   &  6.81 & 2.61 &  2.61 &  0.009 & \adi{[1.68,11.9]} \\
Gender (Male)                         & -2.97 & 2.57 & -1.15 &  0.249 & \adi{[-8.04,2.09]} \\
Chatbot use frequency                  &  2.57 & 0.58 &  4.42 & <0.001 & \adi{[1.43,3.72]}\\
Task success      &  2.30 & 1.99 &  1.16 &  0.248 & \adi{[-1.61,6.22]} \\
Style (Fr) $\times$ Sex (M) & -2.87 & 3.65 & -0.79 &  0.432 & \adi{[-10.1,4.31]} \\
\bottomrule
\end{tabular}
\caption{Linear regression predicting ICSI satisfaction scores from communication style, gender, their interaction, chatbot-use frequency, and task success.}
\label{tab:icsi_sexAndFrequency}
\end{table}

Chatbot-use frequency was also a strong positive predictor of satisfaction (b = 2.57, SE = 0.58, t = 4.42, p < 0.001; \adi{standardized $\beta$ = 0.24, 95\% CI [0.13, 0.35]}), indicating that more frequent chatbot users tended to evaluate the interaction more favorably. In contrast, neither participant gender nor task success significantly predicted ICSI scores.

\dk{As with task success, the style × gender interaction was not statistically significant. We therefore do not interpret gender as a moderator of the satisfaction effect.} \nuria{We conducted exploratory simple-effects contrasts using estimated marginal means to describe the estimated effects within each gender. Among women, \naviw$_f$ was associated with significantly higher satisfaction than \naviw$_d$ (estimate = -6.81, SE = 2.61, t(306) = -2.61, $p < 0.01$), corresponding to an average increase of approximately 7 points on the ICSI scale. Among men, the estimated difference was smaller and did not reach statistical significance (estimate = -3.94, SE = 2.57, t(306) = -1.54, $p > 0.1$), although the estimated effect was in the same direction. These subgroup contrasts provide a descriptive summary of the estimated satisfaction differences within each gender. However, because the overall communication style x gender interaction was not statistically significant, they should not be interpreted as evidence that communication style affects satisfaction differently for women and men. Instead, they suggest a pattern worth investigating in future studies with greater statistical power.}

\definecolor{mygray}{HTML}{E5E5E5}
\begin{tcolorbox}[breakable,colback=mygray!30]
Higher chatbot use frequency was associated with a reduced likelihood of success but higher satisfaction with the interaction. The effect of communication style did not differ between men and women. 
\end{tcolorbox}

\subsection{RQ3: Linguistic accommodation}
\label{sec:results.rq3}
In addition to examining effects on task success and satisfaction, we assessed whether the participants' communication behaviors varied as a function of the chatbot they interacted with. \dk{We first report the average number of turns exchanged during the interaction and the average length of the participants' messages as contextual indicators of task engagement. These measures are not, by themselves, direct evidence of linguistic accommodation, but they help determine whether any accommodation effects could be explained by broader differences in interaction volume. We then examine linguistic accommodation more directly through surface-level style measures and LIWC-based grammatical and semantic features.}

\subsubsection{\dk{Interaction volume as contextual evidence}}

\begin{table}[ht]
\centering
\begin{tabular}{lccccc}
\toprule
\textbf{Predictor} & \textbf{Estimate} & \textbf{SE} & \textbf{t} & \textbf{p} &\textbf{\adi{95\% CI}} \\
\midrule
Intercept                           & 11.86 & 0.84 & 14.06 & <0.001 & \adi{[10.2,13.5]} \\
Style (Friendly)                    &  0.52 & 0.68 &  0.77 & 0.441 & \adi{[-0.81,1.86]}\\
Gender (Male)                          & -0.29 & 0.67 & -0.43 & 0.666 & \adi{[-1.61,1.03]}\\
Chatbot use frequency                   &  0.03 & 0.15 &  0.18 & 0.856 & \adi{[-0.27,0.33]}\\
Task success      &  1.83 & 0.52 &  3.54 & <0.001 & \adi{[0.81,2.85]}\\
Style (Fr) $\times$ Sex (M) & -0.49 & 0.95 & -0.52 & 0.604 & \adi{[-2.36,1.37]}\\
\bottomrule
\end{tabular}
\caption{Linear regression predicting the number of chat messages from communication style, gender, their interaction, chatbot-use frequency, and task success.}
\label{tab:numMessages_sexAndFrequency}
\end{table}

To assess whether the manner in which participants communicated with the system differed across chatbot conditions, we analyzed both the number of chat messages exchanged and the average length of messages participants produced. First, a linear regression predicting message counts from communication style, participant gender, their interaction, chatbot use frequency, and task success revealed a significant overall model, F(5, 306) = 3.13, p < 0.01 (adjusted $R^2$ = 0.033). Neither communication style nor gender predicted how many messages participants sent 
and the style × \dk{gender} interaction was also non-significant.
Chatbot use frequency likewise showed no relationship with message volume.
The only significant predictor was task success: participants who successfully completed the navigation task sent, on average, 1.83 more messages than those who were unsuccessful (b = 1.83, SE = 0.52, t = 3.54, p < 0.001; \adi{Cohen's d = 0.45, 95\% CI [0.21,0.69]}), suggesting that higher message counts primarily reflected deeper engagement with the task rather than differences attributable to style, gender, or chatbot use frequency.

A complementary analysis examined whether the length of the participants’ messages varied across the same predictors. Across all models, no significant effects of communication style, participant gender, chatbot use frequency, task success, or the style × gender interaction emerged for average message length.
This indicates that while successful participants tended to send more messages overall, participants did not adjust the verbosity or phrasing of their individual messages in response to the chatbot’s communication style or their own demographic or performance characteristics. 

In sum, these analyses suggest that \dk{participants' interaction volume} remained largely stable across chatbot conditions, with the number of conversational turns -- but not average message length -- varying as a function of task engagement rather than style-related \dk{adaptation. This provides a contextual baseline for the linguistic accommodation analyses reported below.}

\subsubsection{Linguistic accommodation}

To study the existence of linguistic accommodation we first examined surface-level measures of potential linguistic accommodation of the participants to \naviw's communication style, such as message length, stylistic similarities, and lexical overlap. Second, we conducted a more fine-grained linguistic analysis using Linguistic Inquiry and Word Count (LIWC) \cite{tausczik2010psychological}, which allowed us to capture grammatical and semantic convergence beyond surface metrics. %

\paragraph{Surface-level analysis}
We investigated to which degree the conversation length, language style and word frequencies differed when participants interacted with \naviw$_f$ vs \naviw$_d$. 

Regarding \emph{conversation length}, even though \naviw$_f$ generated more words per turn than \naviw$_d$ \nuria{(consistent with the friendly communication style)}, we did not identify any statistically significant difference in the number of conversational turns and the mean length (in number of words) of the messages exchanged with the different versions of \navi as indicated in our analysis of RQ1 and RQ2. These results suggest an absence of linguistic accommodation from the perspective of message length. 

With respect to \emph{stylistic} adaptation, we applied the sentence style embeddings proposed by Wegmann et al. \cite{Wegmann2022}, and found no evidence that the participants' styles were closer to that of the version of \navi they interacted with. It remains unclear whether this outcome reflects limitations of the embeddings in capturing stylistic variation or a genuine lack of style convergence. Nonetheless, the overall conclusion is that no evidence of stylistic accommodation was observed. Results of the tests conducted here can be found in \ref{sec:appendix.surfaceLevelTests}.

Finally, concerning \emph{positively valenced word frequencies}, \naviw$_f$ employed a slightly different vocabulary, incorporating more positively valenced words than \naviw$_d$. To test whether participants adapted to this vocabulary, we extracted word vectors based on the most positive sentiment words and examined their usage overlap between \navi and participants. While the analysis confirmed that \naviw$_f$ consistently displayed greater positivity than \naviw$_d$, no clear evidence of lexical accommodation was found.   

In summary, although the two versions of \navi differed in linguistic expression, these differences did not appear to elicit corresponding changes in participants' communicative behavior. This absence of accommodation may be attributable to the highly constrained task environment, the well-defined nature of the task, or the relatively short length of the interactions. \ref{sec:appendix.surfaceLevelTests} provides a detailed description of these analyses. The broader implications of these findings are discussed in Section~\ref{sec:discussion}.

\paragraph{LIWC-based analysis}

To complement the previously described surface-level analysis, we conducted a targeted linguistic analysis using LIWC to examine whether more fine-grained grammatical and semantic features revealed signs of convergence between \navi and the participants. 

Statistical relationships between words were assessed using Spearman's rank-order correlation coefficients ($\rho$), with $\alpha$ set at $0.05$ as in the previous analyses. This non-parametric approach allowed us to capture monotonic associations between \naviw's and the participants' linguistic profiles while minimizing sensitivity to outliers. Prominent LIWC categories included grammatical markers (\emph{e.g.}, pronouns, tense usage, numerical references) and semantic domains (\emph{e.g.}, affective language, social references, work-related terms)\dk{. Since this analysis involved a large number of category-level correlations tests, we controlled the false discovery rate (FDR) using the Benjamini-Hochberg procedure \cite{benjamini1995controlling}.  The correction was applied within each analytic setting (aggregated, friendly-only, direct-only) across all 37 LIWC categories that met the variability criterion for analysis (16 grammatical and 21 semantic categories). We report uncorrected p-values together with FDR-adjusted q-values, and we interpret a correlation as statistically reliable only if it remained significant after the correction (q < 0.05). Correlations that did not survive the correction are treated as descriptive patterns only, in line with the exploratory framing of this analysis.}

\dk{Across the aggregated dataset,} small to moderate positive correlations \dk{that survived the FDR correction} were observed between \naviw's and the participants' utterances, particularly in the use of pronouns ($\rho$ = 0.249, p < 0.001, \dk{q < 0.001}), personal pronouns ($\rho$ = 0.306, p < 0.001, \dk{q < 0.001}), first-person singular forms ($\rho$ = 0.290, p < 0.001, \dk{q < 0.001}), verbs ($\rho$ = 0.261, p < 0.001, \dk{q < 0.001}), and past tense ($\rho$ = 0.297, p < 0.001, \dk{q < 0.001}). Numerical expressions showed the strongest effect ($\rho$ = 0.386, p < 0.001, \dk{q < 0.001}), suggesting that participants tended to mirror \naviw's use of quantitative references. \dk{Conjunctions, by contrast, showed a negative association ($\rho$ = -0.228, p < 0.001, \dk{q < 0.001}), indicating divergence rather than convergence for this category.}

When broken down by style, \dk{the two conditions showed different patterns of grammatical alignment.}
\naviw$_d$ displayed stronger alignment overall. \dk{Correlations that remained significant after FDR correction included personal pronouns ($\rho$ = 0.457, p < 0.001, q < 0.001), first-person singular forms ($\rho$ = 0.296, p < 0.001, q < 0.01), auxiliary verbs ($\rho$ = 0.281, p < 0.001, q < 0.01), verbs ($\rho$ = 0.271, p < 0.001, q < 0.01), pronouns ($\rho$ = 0.252, p < 0.01, q < 0.01), numbers ($\rho$ = 0.245, p < 0.01, q < 0.01), and second-person forms ($\rho$ = 0.204, p < 0.05, q < 0.05). In the friendly condition, fewer grammatical correlations survived the correction, namely numbers ($\rho$ = 0.357, p < 0.001, q < 0.001), and past tense ($\rho$ = 0.241, p < 0.01, q < 0.05).}

The analysis of the semantic categories further supported these findings. Alignment in the use of affective words was weak but with significant correlations in the aggregated dataset ($\rho$ = 0.145, p < 0.01, \dk{q < 0.05}) and stronger effects in both the direct ($\rho$ = 0.415, p < 0.001, \dk{q < 0.001}) and friendly ($\rho$ = 0.237, p < 0.01, \dk{q < 0.05}) conditions. Positive emotion words accounted for most of this effect (aggregated: $\rho$ = 0.142, p < 0.05, \dk{q < 0.05}; direct: $\rho$ = 0.406, p < 0.001, \dk{q < 0.001}; friendly: $\rho$ = 0.263, p < 0.001, \dk{q < 0.01}), while negative emotion words \dk{survived the correction only} in the friendly condition ($\rho$ = 0.231, p < 0.01, \dk{q < 0.05}) and, weakly, in the aggregated setting ($\rho$ = 0.156, p < 0.01, \dk{q < 0.05}). The strongest convergence across all analyses was found for work-related vocabulary in the aggregated ($\rho$ = 0.425, p < 0.001, \dk{q < 0.001}) and direct ($\rho$ = 0.564, p < 0.001, \dk{q < 0.001}) settings. \dk{In the friendly condition, the correlation did not survive the correction} ($\rho$ = 0.157, p < 0.05, \dk{q > 0.1}). \dk{Additional semantic categories that remained significant after the correction included social processes (direct: $\rho$ = 0.251, p < 0.01, q < 0.01), relativity (aggregated: $\rho$ = 0.211, p < 0.001, q < 0.001; direct: $\rho$ = 0.212, p < 0.01, q < 0.05), and, in the friendly condition, visual perception ($\rho$ = 0.259, p < 0.01, q < 0.01) and spatial references ($\rho$ = 0.276, p < 0.001, q < 0.01). Overall, 15 of the 37 tested categories survived the FDR correction in the aggregated dataset, 13 in the direct condition, and 7 in the friendly condition.}

These results provide a \dk{cautious and} nuanced picture of linguistic \dk{accommodation. Participants} did not exhibit broad stylistic or lexical accommodation, \dk{but} selected grammatical and semantic \dk{categories showed associations between \naviw's and participants' language. We interpret these patterns as feature-level alignment} rather than \dk{as evidence of} a complete shift in \dk{users'} communication style. Complete descriptive statistics and detailed results are provided in ~\ref{sec:appendix.liwc}.

Given the number of models and language-related analyses conducted, the results should be interpreted with an emphasis on consistent patterns across outcomes rather than isolated p-values.

\definecolor{mygray}{HTML}{E5E5E5}
\begin{tcolorbox}[breakable,colback=mygray!30]
In sum, neither communication style, gender nor chatbot use frequency predicted the number of messages sent by the participants. The only significant predictor was task success: participants who successfully completed the navigation task sent more messages than those who were unsuccessful. Moreover, participants did not exhibit broad stylistic or lexical accommodation. \dk{The LIWC results suggest selective feature-level alignment in specific} grammatical and semantic domains, \dk{which should be interpreted cautiously given the constrained task and the number of language-related analyses}.
\end{tcolorbox}

\section{Discussion}
\label{sec:discussion}
In this section, we elaborate on several interrelated themes emerging from our study that provide a framework for interpreting the findings and contextualizing them within broader debates on human-chatbot interaction. Specifically, we interpret the findings in relation to RQ1 to RQ3 and focus on (i) how fixed stylistic tone relates to subjective satisfaction and task success; (ii) why the control condition outperformed both chatbot variants in this specific task setting, and (iii) what the observed, feature-specific linguistic accommodation suggests about short, goal-oriented interactions. 

\subsection{Revisiting gendered navigation differences}
Our results reveal no significant gender differences in task performance under the control condition (without any chatbot), which is noteworthy given previous research that has reported gender differences in navigation tasks, with men relying more on metric strategies (\emph{e.g.,} cardinal directions) and women favoring landmark-based or route strategies \cite{lawton1994gender,coluccia2004gender}. More recent research, however, suggests that such differences are highly context-dependent and can be moderated by task design, environmental cues, and motivational factors \cite{nazareth2019meta}. Recent studies further indicate that gender gaps may shrink under specific motivational or environmental conditions: for example, Schinazi et al. \cite{schinazi2023motivation} showed that time pressure can eliminate female disadvantages in virtual navigation tasks, while Dahmani et al. \cite{dahmani2023sex} found that environments offering both landmark and metric cues result in comparable performance between men and women. The absence of a gender effect in our control condition therefore reflects a strength of the task design, which provided a balanced environment that did not favor one navigational strategy over another. Thus, the control condition can be seen as an example of good design practice: it allowed participants to draw on their respective strengths while minimizing the impact of their weaknesses, establishing a task-specific neutral baseline against which the effects of \naviw's communication styles can be meaningfully evaluated. At the same time, this “neutral baseline” should be interpreted as a property of our particular 2D task: because the map was visible and visually informative, some participants may have relied on map cues (\emph{e.g.,} icons, layout, and spatial constraints) to infer the target location with relatively limited need for dialogue. This means we should avoid generalizing the control-condition pattern to real-world navigation or to tasks where the environment is not continuously visible or where the information is genuinely unavailable without guidance. Crucially, the central gender-related signal in our study does not concern baseline navigation performance, but rather the interaction between user characteristics and stylistic framing in the chatbot conditions.

\subsection{The limits of chatbots and techno-solutionism}

Within the chatbot conditions, the direct style most closely approximated the control condition. However, participants preferred and performed better with the friendly version of \navi than with the direct one, yet neither outperformed participants in the control condition, suggesting that the introduction of a conversational agent in the task at hand changes the dynamics in ways that are not necessarily beneficial for objective task performance, even when subjective evaluations of the interaction are more positive. One plausible explanation is interaction overhead: in a visually informative, low-ambiguity map task, formulating queries and processing conversational responses may impose additional cognitive and temporal costs compared to following stepwise written instructions. Prior work has reported that users often approach chatbots with different social expectations when compared to non-agent based systems \cite{nass2000machines,luger2016like}, which might partially explain this finding. Interestingly, scholars have suggested that conversational interfaces are beneficial over direct manipulation when the task complexity is greatest, proposing navigation paths as one of the use cases where dialogue interfaces should have an advantage \cite{brennan1990conversation}. However, we do not find any supporting empirical evidence of this claim in the 2D fictional navigation task of our user study.

The superior performance of participants in the control condition when compared to using \navi underscores the need for caution in assuming that chatbots are universally helpful, particularly in tasks where information is already available through a well-designed non-conversational interface. As previous critiques of techno-solutionism argue \cite{morozov2013save,shneiderman2020human}, technological interventions should be carefully evaluated not only for the effectiveness but also for broader considerations, including environmental cost, data privacy and appropriateness for the task at hand. In our study, an additional practical factor is that participants were instructed to minimize conversational turns which may have constrained natural interaction and may also blur interpretation when conversational turns or related interaction indicators are discussed as outcomes. We therefore interpret the control condition's advantage conservatively as evidence that, in a visually rich 2D map task, adding a chatbot layer can introduce interaction overhead without clear performance gains, rather than as a general claim that chatbots are inferior for navigation overall. Chatbots may lead to higher levels of engagement and persistence in some contexts, but they are not a substitute for all forms of human-computer interaction, as illustrated by our findings. More precisely for this paper, our data support a narrower conclusion: chatbots are not universally helpful in structured, map-visible tasks where direct interaction already provides efficient access to relevant cues.

\subsection{Communication style matters for task success}

Task success, measured as the correct identification of the embassy’s location, was significantly higher in the friendly than in the direct condition, indicating that stylistic tone can influence performance when users rely on conversational guidance. A plausible explanation is that the friendly framing provided more supportive, relational cues that encouraged continued engagement with the task (\emph{e.g.,} sustained interaction, more opportunities to request clarification, and greater time spent working through the steps), which can be beneficial in multi-step guidance scenarios. Consistent with this hypothesis, participants preferred interacting with \naviw$_f$. Furthermore, they engaged in more conversational turns and typed more words in their interactions with \navi when they were successful at the task, irrespective of the condition, which could have increased their exposure to \naviw’s communication style. 

At the same time, our results indicate that prior experience with chatbots systematically shaped user outcomes: higher chatbot-use frequency was associated with greater subjective satisfaction but a lower likelihood of task success. This dissociation is consistent with prior work showing that repeated and familiar interactions with conversational agents can enhance perceived quality, comfort, and relational evaluations without necessarily improving instrumental performance \cite{Araujo2024}. In parallel, frequent users may be more prone to over-reliance on automated guidance, relying on the agent’s recommendations while engaging less in active verification against available task cues, a pattern known to increase the risk of accuracy costs in structured, low-ambiguity tasks \cite{Skitka2000,Parasuraman2010}.

Our findings also resonate with recent evidence that the communication styles in conversational agents strongly influence user perceptions and behaviors. For instance, Poivet et al. demonstrated in a narrative detective game that cooperative communication styles were associated with higher perceived warmth, greater user preference, and more engaging conversations, whereas aggressive styles elicited longer inputs and were more frequently associated with suspicion and negative judgments \cite{poivet2023influence}. 

These results support a careful interpretation: a friendly style can improve subjective experience and may support task success in guidance-oriented contexts, while overly direct styles may reduce engagement or perceived support.

While we did not identify evidence of linguistic accommodation in our surface-level analysis, the larger number of turns among successful participants could have enabled them to differentiate better between communication styles and benefit from it. This finding aligns with theories of engagement, persistence and motivation in HCI research \cite{ryan2000self,dmello2017advanced}, suggesting that relational cues, even if subtle, may scaffold user persistence in demanding tasks. It is also concordant with research in motivational psychology as social support and encouragement have been found to enhance self-efficacy and perseverance in problem-solving contexts \cite{lakey2000social}. Regarding conversational agents, social presence theory has reported that perceived warmth and responsiveness can increase trust and willingness to interact \cite{bickmore2005establishing}, which in turn may contribute to better outcomes in task success. Because our empirical measures are task success, satisfaction, and interaction behavior (\emph{e.g.,} turns, help requests, and time on task), we frame this point in terms of sustained engagement and task-relevant interaction rather than unmeasured constructs.

Another interpretation of this finding is that a friendly communication style may reduce the cognitive load in the interaction by framing the task as more collaborative than transactional. According to prior work on politeness and dialogue systems, supported feedback and mitigated directives can enhance the users’ willingness to continue interacting, even when facing obstacles \cite{ribino2023role}.

\subsection{Gender-related patterns are exploratory}
\nuria{We did not identify a significant interaction between communication style and gender neither for task success nor for satisfaction. Exploratory gender-stratified patterns were directionally consistent with a stronger benefit of the friendly communication style among female participants. This finding is theoretically interpretable in light of prior work.}
Previous research has reported that women are often more responsive to relational cues –- such as friendliness and encouragement -– than men \cite{tannen1990you}. Research in social psychology also suggests that women may have a greater ability to recognize and benefit from supportive communication styles \cite{tenenbaum2011telling}. Similar preferences for cooperative communication styles have also been documented more generally across user groups, with participants reporting greater warmth, engagement, and conversational preference when interacting with cooperative rather than aggressive agents \cite{poivet2023influence}. \nuria{Nevertheless, given the lack of significance in the interaction, the gender-stratified simple effects are better understood as hypothesis-generating patterns that may inform future studies with designs specifically powered and preregistered to test such moderation.}

\nuria{This distinction matters for chatbot design. Our findings do not justify simple gender-based personalization rules, such as automatically assigning a friendly communication style to women and a direct style to men. Such rules would risk overgeneralizing from subgroup patterns and reinforcing stereotypes. A more defensible implication is that users may differ in their preferred balance between relational support and concise task focus, and that future systems should offer transparent, user-controllable style options rathen than infer style preferences from demographic categories.}

\subsection{The nuanced picture of linguistic accommodation}
\nuria{RQ3 examined whether chatbot style affects the users' own language during the interaction. This question is practically important because accommodation would imply that communication style not only impacts how the interaction is evaluated afterwards but also changes the interaction dynamics. Our results provide limited evidence for such an effect.} Surface-level measures –- such as message length, stylistic embeddings and positively valanced vocabulary overlap -– did not reveal clear signs of accommodation \nuria{to the chatbot's communication style. The participants' language did not generally become more friendly, more direct, more verbose or more stylistically similar to the version of \navi they interacted with.} 

The LIWC-based analyses uncovered selective but systematic \nuria{feature-level alignment in some grammatical and semantic categories, including} pronouns, temporal references, numerical expressions and work-related vocabulary. \nuria{These patterns are compatible with} Communication Accommodation Theory, which highlight that \nuria{accommodation can be} selective \nuria{rather than global} in human–machine interaction and propose an expanded framework for accommodation in technologically mediated contexts \cite{giles2023communication}. \nuria{However, they should be interpreted with caution. In a constrained navigation task, numerical references, route-related terms and temporal expressions may reflect the structure of the task itself as much as interpersonal convergence. Given the number of language-related analyses, isolated significant correlations should not be treated as strong evidence of broad accommodation. The more conservative conclusion is that short, goal-oriented chatbot interactions may produce limited and feature-specific alignment instead of a general adoption of the chatbot's communication style.}

\dk{The boundary condition is valuable for design as it suggests that adjusting a chatbot's style may influence perceived communication quality without necessarily changing how users formulate their own messages. Designers should therefore avoid assuming that a friendly or a direct agent will automatically induce its users to use a similar communicative style during brief task-focused exchanges. We leave to future work the exploration of language accommodation in longer interactions.}

The interpretation of these results is closely tied to the nature of the task. The interaction with \navi consisted of short, goal-oriented exchanges with relatively limited utterances, which inherently constrained the scope for broader stylistic adaptation. The observed convergence in selected linguistic features should therefore be understood as adaptation within the frame of the communication that took place within our navigation-oriented task rather than as evidence of global style alignment.

\subsection{Implications for chatbot design}
From our findings we derive several implications for the design of conversational agents. 

First, designers should critically assess whether a chatbot is the most appropriate interaction paradigm for their intended application. While conversational agents can offer natural and engaging forms of interaction, they are not universally the best fit. In some contexts, more direct interaction modalities -- such as those in our control study -- may provide greater efficiency, clarity and accessibility. Careful consideration of user needs, task demands and contextual factors is essential before committing to a chatbot-based solution. 

Second, \nuria{communication style should be treated as a consequential but context-sensitive design parameter. A friendly communication style can improved perceived communication quality and may support task success in guidance-oriented interactions, such as the task described in this paper. However, the size and practical value of such effects should be evaluated against the task context and alternative interface designs.}

Third, cultural variation should also be taken into account: expectations of directness, politeness and friendliness vary across contexts and cultures \cite{hofstede1984culture}, which could shape the effectiveness of chatbot communication strategies. While we did not explicitly study cultural variation in our study, we plan to do so in future work.  

\nuria{Fourth, style personalization should be transparent and preference-based rather than demographically deterministic. The exploratory gender-related patterns identified in this study point to possible variation in the users' sensitivity to relational cues, but do not support automatic gender-based style assignment.} Thus, transparency and user control should be prioritized, allowing users to adjust or override the chatbot’s communication style to mitigate a misalignment between system defaults and individual preferences.

\dk{Fifth, designers should not assume that users will linguistically adapt to an agent's communication style in short interactions. If accommodation is a design goal, for example in learning, coaching or therapeutic support tasks, it should be measured directly and studied over longer interactions.}

Finally, \dk{communication-style decisions in} chatbot design should be \dk{evaluated inclusively and cross-culturally. Expectations about directness, politeness and friendliness vary across contexts and cultures \cite{hofstede1984culture}. Systems optimized for one population or task type may not generalize to others.} In addition, where conversational turns or related interaction indicators are used as outcome measures, study procedures should avoid instructions that directly constrain those indicators (or should explicitly model such constraints), to keep interpretations of ``efficiency'' and ``engagement'' clearly separated from procedural compliance.

\section{Limitations and future work}
\label{sec:limitations}

Our study is not without limitations. First, \nuria{our findings correspond to a Prolific sample of English-speaking, neurotypical participants based in the USA, performing a short fictional map-based navigation task in a controlled setting. Hence, they might not generalize to different populations or tasks. Second,} our analysis treated gender as a binary variable failing to capture the diversity of gender identities. Specifically, we used the self-reported ``sex'' variable that participants entered on Prolific as a response to the question \emph{``What is your sex, as recorded on legal/official documents?''}. 
\nuria{Third}, although timing data was collected, we did not use it as a reliable measure of task efficiency because delays in task completion could have been due to a variety of reasons outside of the scope of the task, such as distractions, network issues or multitasking. \nuria{Fourth}, our 2D navigation task while realistic corresponded to a fictional situation rather than a real-world environment, \dk{where environmental uncertainty, stress, mobility, device use, sensory cues and consequences of error may be substantially different. The control's condition advantage should therefore be interpreted as specific to a visually available, structured map task rather than as evidence against chatbots for navigation more generally.} \dk{Fifth, the interaction was short and one-off. Longer-term use may produce different patterns of satisfaction, trust calibration, reliance and linguistic accommodation. In particular, communication style adaptation may require repeated exposure, richer interactional history or higher social stakes than were present in this study.}
\nuria{Sixth, the subjective measure of communication satisfaction was adapted from the ICSI \cite{hecht1978conceptualization}. Although this provided a theoretically relevant measure of perceived communication quality, some items may be less natural in short human-chatbot interactions than in interpersonal communication. Future work should compare the full scale with task-specific or reduced item sets and report whether the pattern of results remains stable. Finally, several statistically significant effects explained only a modest proportion of variance, and the linguistic analyses involved multiple feature-level tests. These results are useful for identifying patterns, but they should be interpreted with emphasis on convergent evidence and practical magnitude rather than isolated p-values.}
 
\dk{Future research should} replicate the study with participants from a significantly different culture to investigate cross-cultural differences in communication expectations, and examine \nuria{longitudinal} interactions where style adaptation and user learning may play a greater role. Furthermore, we would like to extend the study to a task of a different nature to shed light on the generalizability of our findings beyond the current goal-oriented, multi-step scenario. \nuria{Finally, future research could validate communication satisfaction measures specifically for conversational AI or develop context-specific instruments that better capture user experiences in goal-oriented chatbot interactions while retaining strong psychometric properties.}

\section{Conclusion} \label{sec:conclusion}
In this paper, we have studied how chatbot communication style influences task performance, user satisfaction and linguistic accommodation in a fictional 2D navigation task by means of a user study where participants interacted with two stylistic variants of a navigation chatbot, \naviw, \dk{or completed the same task using  static step-by-step instructions without chatbot interaction}. Within the chatbot conditions, the friendly communication style increased users' satisfaction \dk{and was associated with higher task success than the direct style. We do not find statistically significant evidence that these effects differed by gender, although exploratory subgroup analyses suggested patterns that may warrant investigation in future work.} We also found only partial evidence of linguistic accommodation, indicating selective alignment on individual linguistic features rather than a broad accommodation effect. Furthermore, participants in the control condition achieved higher task success than participants interacting with either chatbot, which raises questions regarding the suitability of chatbots for certain tasks, despite the benefits of a more supportive conversational style. \dk{Our findings support a cautious design principle: communication style matters, but it should be selected and evaluated in relation to task demands, user preferences, transparency and practical effectiveness.}

\section*{Acknowledgments}
This work has been supported by a nominal grant received at the ELLIS Unit Alicante Foundation from the Regional Government of Valencia in Spain (Resolución de la Conselleria de Industria, Turismo, Innovación y Comercio, Dirección General de Innovación) and has received funding from the European Union’s Horizon Europe research and innovation programme under Grant Agreement No: 101120237 (ELIAS). E.D. and A.G. have been supported by Banco Sabadell Foundation. E.D., A.G. and N.O. have been supported by Intel  Corporation. R.A.B. has been supported by the Applied Data Science Focus Area from Utrecht University, the Netherlands. E.D.’s work was co-funded by the European Union under the project ROBOPROX (reg. no. CZ.02.01.01/00/22\_008/0004590).

\paragraph{Disclaimer} Funded by the European Union. Views and opinions expressed are however those of the author(s) only and do not necessarily reflect those of the European Union or European Commission. Neither the European Union nor the granting authority can be held responsible for them. Grant agreement no. 101120237

\newpage

\appendix

\section{Navigation scenario setup}\label{appendix:system_prompts}

This appendix presents the system prompts used to instruct \navi --based on ChatGPT (\texttt{gpt-4.1-2025-04-14})-- with the navigation scenarios and assign the friendly or direct persona to \naviw$_f$ or \naviw$_d$, respectively. Furthermore, it lists the navigation instructions in the control condition.

\subsection{Friendly \navi}
\label{sec:appendix.npsi.friendly}

The following system prompt was used to control the behavior of \naviw$_f$:

\definecolor{mygreen}{HTML}{99C000}
\begin{tcolorbox}[breakable,colback=mygreen!30]
\noindent You are NAVI, a helpful and approachable citizen of a given city. You are at the central train station. Your objective is to give instructions through the city to the embassy to a tourist who lost their phone and documents while ensuring the instructions are given gradually and conversationally, fostering a multi-turn dialogue. Always adhere strictly to the given scenario for directions:
\begin{itemize}
    \item You and the tourist needing help are both at the central train station.
    \item From the central train station, go four blocks to the west. There is a church there.
    \item Only when asked, identify the correct church, which is the tallest one.
    \item From the church, walk north until you spot the town hall, a tall 6-storey building.
    \item Only when asked, distinguish the town hall from a neighboring hotel by the large coat of arms of the city on it.
    \item Turn east. At the end of the street, you will see the castle, which is near the embassy.
    \item Only when asked, specify that the correct castle has four towers, not two.
    \item The embassy is to the west of the castle.
    \item Only when asked, state that you cannot miss it because it is across the street from the post office.
\end{itemize}
Refrain from giving all the instructions at once, encouraging the tourist to ask clarifying questions or confirm details. The tourist should ask for clarification themselves when there are more similar landmarks. Do not give this information away yourself immediately, but strictly only if asked. If the tourist gets confused, patiently reiterate the instructions from the beginning and ensure understanding. Your responses must be warm, helpful, and polite, creating an authentic conversational experience. Give local tips when appropriate. If you are asked for anything that is not explicitly stated in the scenario, always steer the conversation to the scenario. Do not provide any details or clarifications except for the ones clearly stated above.
\end{tcolorbox}

\subsection{Direct \navi}
\label{sec:appendix.npsi.direct}

The following system prompt was used to control the behavior of \naviw$_d$:

\definecolor{mypurple}{HTML}{c55a92}
\begin{tcolorbox}[breakable,colback=mypurple!20]
\noindent You are NAVI, a citizen of a given city, and you communicate directly with people. You are at the central train station. Your objective is to give instructions through the city to the embassy to a tourist who lost their phone and documents while ensuring the instructions are given gradually and conversationally, fostering a multi-turn dialogue. Always adhere strictly to the given scenario for directions:
\begin{itemize}
    \item You and the tourist needing help are both at the central train station.
    \item From the central train station, go four blocks to the west. There is a church there.
    \item Only when asked, identify the correct church, which is the tallest one.
    \item From the church, walk north until you spot the town hall, a tall 6-storey building.
    \item Only when asked, distinguish the town hall from a neighboring hotel by the large coat of arms of the city on it.
    \item Turn east. At the end of the street, you will see the castle, which is near the embassy.
    \item Only when asked, specify that the correct castle has four towers, not two.
    \item The embassy is to the west of the castle.
    \item Only when asked, state that you cannot miss it because it is across the street from the post office.
\end{itemize}
Refrain from giving all the instructions at once, encouraging the tourist to ask clarifying questions or confirm details. The tourist should ask for clarification themselves when there are more similar landmarks. Do not give this information away yourself immediately, but strictly only if asked. Do not suggest any follow-up or clarification yourself. Your responses must be direct, brief, curt, and to the point. If you are asked for anything that is not explicitly stated in the scenario, always steer the conversation to the scenario.
\end{tcolorbox}

\subsection{Control condition}
\label{sec:appendix.npsi.control}

In the control condition, the following navigation instructions were revealed to the participants on a step-by-step fashion (participants had to press a button to proceed to the next step):
\begin{enumerate}
  \item From the central train station, go four blocks to the west. There is a church there.
  \item The correct church is the tallest one.
  \item From the church, walk north until you spot the town hall, a tall 6-storey building.
  \item You can distinguish the town hall from a neighboring hotel by the large coat of arms of the city on it.
  \item Turn east. At the end of the street, you will see the castle, which is near the embassy.
  \item The correct castle has four towers, not two.
  \item The embassy is to the west of the castle.
  \item You cannot miss it because it is across the street from the post office.
\end{enumerate}

\section{Post-task questions}
\label{sec:appendix.questions}

In this appendix, we list the exact wording of the questions presented to the participants in the post-task survey.

\subsection{ICSI: Communication quality and satisfaction (Block A)}
\label{sec:appendix.questions.a}

The communication quality and satisfaction questionnaire, listed below, was presented to the participants in the chatbot condition with the following instructions: 

\begin{tcolorbox}[breakable,colback=yellow!5!white]
\noindent The purpose of this questionnaire is to investigate your reactions to the conversation you just had. Below, you will be asked to react to a number of statements. Please indicate the degree to which you agree or disagree that each statement describes this conversation. \textbf{Note that this is a standardized communication satisfaction questionnaire, so please do your best to answer the questions even if they may seem irrelevant to the conversation with NAVI in some cases.}
\end{tcolorbox}
\begin{enumerate}
\item NAVI let me know that I was communicating effectively.
\item Nothing was accomplished.
\item I would like to have another conversation like this one.
\item NAVI genuinely wanted to get to know me.
\item I was very dissatisfied with the conversation.
\item I had something else to do.
\item I felt that during the conversation I was able to present myself as I wanted NAVI to view me.
\item NAVI showed me that they understood what I said.
\item I was very satisfied with the conversation.
\item NAVI expressed a lot of interest in what I had to say.
\item I did NOT enjoy the conversation.
\item NAVI did NOT provide support for what they were saying.
\item I felt I could talk about anything with NAVI.
\item We each got to say what we wanted.
\item I felt that we could laugh easily together.
\item The conversation flowed smoothly.
\item NAVI changed the topic when their feelings were brought into the conversation.
\item NAVI frequently said things which added little to the conversation.
\item We talked about something I was NOT interested in.
\end{enumerate}

\subsection{User experience and preferences (Block B) -- chatbot condition}
\label{sec:appendix.questions.b-navi}

The following questions were given to participants in the chatbot condition:
\begin{enumerate}
\item The conversation felt natural and human-like.
\item NAVI effectively helped me solve the task.
\item I would have preferred to have asked a human.
\item How frequently do you use chatbots?
\end{enumerate}

\subsection{User experience and preferences (Block B) -- control condition}
\label{sec:appendix.questions.b-control}

The following questions were given to participants in the control condition:
\begin{enumerate}
\item I am satisfied with how the instructions conveyed the information on how to get to the embassy.
\item I would have preferred receiving the instructions to get to the embassy from a human rather than in written form.
\item I would have preferred to ask a chatbot for help rather than follow written instructions.
\item How frequently do you use chatbots?
\end{enumerate}

\subsection{Scales}
\label{sec:appendix.questions.scales}

All questions listed in the subsections above were answered using 7-point Likert scales. The standard response options were labeled  
`Strongly disagree',
`Disagree',
`Somewhat disagree',
`Neutral',
`Somewhat agree',
`Agree', and
`Strongly agree',
with the exception of the question `How frequently do you use chatbots?', which used the labels
`Never',
`Rarely',
`Occasionally',
`Sometimes',
`Often',
`Usually', and
`All the time'.

\section{Instructions}
\label{sec:appendix.instructions}

In this appendix, we provide the exact instructions given to study participants in all three conditions (friendly \naviw, direct \naviw, and the control condition).

\subsection{Chatbot condition}

The following instructions were given to study participants in the chatbot condition (friendly or direct \naviw):

\begin{tcolorbox}[breakable,colback=yellow!5!white]
\noindent \textbf{NAVI Study: Instructions}
\vspace{6pt}

\noindent \textbf{This study consists of two parts:}
\begin{itemize}
    \item Chat with NAVI,
    \item Questionnaire.
\end{itemize}
\vspace{6pt}

\noindent \textbf{Chat with NAVI}
\vspace{6pt}

\noindent \textbf{Scenario:} Imagine you have arrived at the train station of a capital city where you have never been before. You realize you don't have your phone and documents, so you need to urgently get to the embassy. You meet \textbf{NAVI}, a local citizen, and ask for the way to the embassy.
\vspace{6pt}

\noindent \textbf{Your task is to get directions to the embassy through a chat with NAVI.}
\vspace{6pt}

\noindent \textbf{It's important that you keep chatting and asking questions until you are confident you know which building on the map is the embassy. NAVI might not give you the exact answer right away, and you may need to ask follow-up questions or clarify details to be sure. On average, you'd need to send 6–8 messages to figure out where the embassy is. An absolute minimum of messages you need to send is 3, but it's very unlikely this would be enough for you to determine the location of the embassy.}
\vspace{6pt}

\noindent Once you are confident you know where the embassy is located, in the \textbf{left sidebar}, you'll click the button \textbf{I know where the embassy is!} Numbers will appear over each building on the map. You will then choose the number of the building where the embassy is. \textbf{Once you confirm, you will move on to the questionnaire.}
\vspace{6pt}

\noindent \textbf{There is a strict limit of 8 minutes for the chat.} The timer starts once you proceed to the chat page. When 3 minutes or less are left, the remaining time will appear in the top-left corner. The timer turns red in the last minute. Once the time is up, the chat is disabled and you have to select the embassy. The remaining time displayed only updates when you send a message, not continuously. \textbf{Please note that having a conversation with NAVI is needed to proceed with the study, since the questionnaire that follows after the chat is focused on your conversation with NAVI.}
\vspace{6pt}

\noindent \textit{Tip: Don't rush! NAVI may not tell you directly. Make sure you have enough information from the chat to identify the correct building on the map, and ask follow-up questions.}
\vspace{6pt}

\noindent The following images explain the chat interface and the map:
\vspace{6pt}

\noindent (see Figure~\ref{fig:interface-navi} and the map description in Figure~\ref{fig:interface-static})
\vspace{6pt}

\noindent \textbf{Questionnaire}
\vspace{6pt}

\noindent The questionnaire consists of two blocks with a total of 25 questions as a follow-up to the chat with NAVI. For each question, there are 7 options to choose from (Likert scale). All questions need to be answered to continue to the next block and to complete the study.
\vspace{6pt}
\end{tcolorbox}

\subsection{Control condition}

The following instructions were given to study participants in the control condition (no chatbot):

\begin{tcolorbox}[breakable,colback=yellow!5!white]
\noindent \textbf{NAVI Study: Instructions}
\vspace{6pt}

\noindent\textbf{Navigation task}
\vspace{6pt}

\noindent \textbf{Scenario:} Imagine you have arrived at the train station of a capital city where you have never been before. You realize you don't have your phone and documents, so you need to urgently get to the embassy. You meet \textbf{NAVI}, a local citizen, who provides you with step-by-step instructions on how to get to the embassy.
\vspace{6pt}

\noindent The following image explains the navigation task interface:
\vspace{6pt}

\noindent (see Figure~\ref{fig:interface-static})
\vspace{6pt}

\noindent \textbf{Your task is to get directions to the embassy by carefully following the instructions given by NAVI. You will be given the instructions one by one, showing the next instruction by clicking the `Show next instruction' button.}
\vspace{6pt}

\noindent Please note that the map image is static, so you will need to keep track of your position in the city yourself.
\vspace{6pt}

\noindent \textbf{Once you reveal all the instructions, you will be asked to choose the building number where you think the embassy is.}
\vspace{6pt}

\noindent \textbf{Finally, you will fill out a short questionnaire to reflect on your experience.}
\end{tcolorbox}

\section{\adi{Impact of Communication Style on User Preferences}}
\label{sec:appendix.impactCommunicationStyle}

\adi{We report the full statistical comparison between the friendly and direct chatbot conditions on the three Block~B outcome measures: perceived naturalness, task helpfulness and human preference. Communication style had no statistically significant impact on these measures. The results are reported below for completeness.}

\adi{As throughout Section~\ref{sec:results}, normality within each condition was assessed with the Kolmogorov-Smirnov test. All three measures departed from normality in both conditions (all KS $p \leq 0.001$), so we report the non-parametric Mann--Whitney U test rather than the independent-samples t-test, whose normality assumption is not met for these variables. Table~\ref{tab:blockb_appendix} summarizes the results.}

\begin{table}[h!]
\centering
\begin{tabular}{lcccc}
\toprule
\textbf{Variable} & \textbf{Friendly} & \textbf{Direct} & \multicolumn{2}{c}{\textbf{Mann--Whitney}} \\
 & \textit{M (Mdn)} & \textit{M (Mdn)} & $W$ & $p$ \\
\midrule
Perceived naturalness   & 5.12 (5) & 4.92 (5) & 11254.00 & 0.238 \\
Task helpfulness        & 6.16 (6) & 5.98 (6) & 11426.50 & 0.318 \\
Human preference        & 3.79 (4) & 3.94 (4) & 12788.50 & 0.427 \\
\bottomrule
\end{tabular}
\caption{\adi{Comparison of the friendly and direct conditions on the Block~B measures (perceived naturalness, task helpfulness and human preference). $M$ and $Mdn$ denote the mean and median, respectively.}}
\label{tab:blockb_appendix}
\end{table}

\section{Surface level linguistic accommodation}
\label{sec:appendix.surfaceLevelTests}

This appendix reports the full set of statistical tests conducted for the surface-level analysis of accommodation effects reported in Section \ref{sec:results.rq3}.

\subsection{Stylistic adaptation}

Table \ref{tab:style_similarity} reports the cosine similarity between the style embeddings of sentences \cite{Wegmann2022} produced by the two versions of \navi (indicated with \_N) and by participants (denoted with \_P) interacting with the respective version. The results show that participant responses were highly similar to one another (as reflected by the high cosine similarities between ``Direct\_P" and ``Friendly\_P"). However, there was little evidence of stylistic alignment between the participants language and the version of \navi with which they interacted (as reflected by the low cosine similarities).

\begin{table}[h!]
\centering
\begin{tabular}{lcccc}
\hline
 & \textbf{Friendly\_N} & \textbf{Friendly\_P} & \textbf{Direct\_N} & \textbf{Direct\_P} \\
\hline
\textbf{Friendly\_N} & 1.0000  & -0.2529 & 0.4246  & -0.2789 \\
\textbf{Friendly\_P} & -0.2529 & 1.0000  & -0.3694 & 0.9941  \\
\textbf{Direct\_N}   & 0.4246  & -0.3694 & 1.0000  & -0.3647 \\
\textbf{Direct\_P}   & -0.2789 & 0.9941  & -0.3647 & 1.0000  \\
\hline
\end{tabular}
\caption{Cosine similarities between the style embeddings of sentences produced by different versions of \navi and participants. The sentences written by participants are denoted by ``\_P" and those generated by \navi are denoted by ``\_N".}
\label{tab:style_similarity}
\end{table}

\subsection{Word frequencies}

Table \ref{tab:word_frequencies} presents the frequency of positively valenced words used by the two versions of \navi and by participants. As expected, \naviw$_f$ employed such words most frequently, supporting the assumption that this version would be perceived as more friendly. However, there was no evidence that participants interacting with \naviw$_f$ adopted a similar tendency. This observation is further confirmed in Table \ref{tab:correlation_matrix}, which reports the cosine similarity of the corresponding word vectors extracted from Table \ref{tab:word_frequencies}.

\begin{table}[ht!]
\centering
\begin{tabular}{lrrrr}
\hline
\textbf{Word} & \textbf{Direct\_N} & \textbf{Direct\_P} & \textbf{Friendly\_N} & \textbf{Friendly\_P} \\
\hline
excellent      & 0  & 0  & 26  & 0 \\
congratulations& 0  & 0  & 2   & 0 \\
fantastic      & 0  & 0  & 15  & 0 \\
lovely         & 0  & 0  & 17  & 0 \\
wonderful      & 0  & 0  & 44  & 0 \\
awesome        & 0  & 0  & 29  & 0 \\
delicious      & 0  & 0  & 2   & 0 \\
amazing        & 0  & 0  & 1   & 0 \\
impressive     & 0  & 0  & 5   & 0 \\
bright         & 0  & 0  & 1   & 0 \\
beautiful      & 0  & 0  & 9   & 0 \\
happy          & 0  & 0  & 112 & 0 \\
glad           & 1  & 0  & 49  & 0 \\
hopefully      & 0  & 0  & 0   & 1 \\
finally        & 2  & 0  & 0   & 0 \\
perfect        & 0  & 2  & 129 & 0 \\
charming       & 0  & 0  & 2   & 0 \\
pretty         & 0  & 0  & 13  & 2 \\
thank's        & 0  & 1  & 0   & 0 \\
enjoy          & 0  & 0  & 25  & 0 \\
pleasant       & 0  & 0  & 3   & 0 \\
welcome        & 44 & 0  & 48  & 0 \\
best           & 0  & 0  & 22  & 1 \\
appreciate     & 0  & 2  & 1   & 0 \\
grateful       & 0  & 1  & 0   & 0 \\
easily         & 1  & 0  & 29  & 1 \\
great          & 8  & 14 & 319 & 7 \\
nice           & 0  & 0  & 40  & 2 \\
fun            & 0  & 0  & 1   & 0 \\
thanks         & 1  & 23 & 5   & 17 \\
progress       & 0  & 0  & 11  & 0 \\
standing       & 0  & 0  & 27  & 2 \\
fastest        & 0  & 1  & 0   & 0 \\
hoping         & 0  & 1  & 0   & 0 \\
bold           & 0  & 0  & 1   & 0 \\
good           & 24 & 3  & 144 & 3 \\
\hline
\end{tabular}
\caption{Frequency of positively valenced words in conversations, comparing the friendly and direct versions of \navi (N) and participants (P) interacting with the respective version.}
\label{tab:word_frequencies}
\end{table}

\begin{table}[h!]
\centering
\begin{tabular}{l|cccc}
\hline
 & \textbf{Direct\_N} & \textbf{Direct\_P} & \textbf{Friendly\_N} & \textbf{Friendly\_P} \\
\hline
\textbf{Direct\_N}    & 1.000 & 0.149 & 0.397 & 0.151 \\
\textbf{Direct\_P}    & 0.149 & 1.000 & 0.475 & 0.958 \\
\textbf{Friendly\_N}  & 0.397 & 0.475 & 1.000 & 0.383 \\
\textbf{Friendly\_P}  & 0.151 & 0.958 & 0.383 & 1.000 \\
\hline
\end{tabular}
\caption{Cosine similarities of the positive valence word vectors for participants and different versions of \naviw.}
\label{tab:correlation_matrix}
\end{table}

\section{LIWC analysis}
\label{sec:appendix.liwc}

In this appendix, we include statistical results from the LIWC-based linguistic analysis. For the analysis, a total of N = 624 text samples were used. Half of the texts represented participant communication (direct: n = 157; friendly: n = 155), while an equal number of texts was obtained from the \navi agent through text extraction (direct: n = 157; friendly: n = 155). We report the LIWC variables that were suitable for statistical analysis. The remaining LIWC variables exhibited insufficient variability and were therefore excluded from the correlation analyses. \dk{To account for multiple comparisons, the false discovery rate (FDR) was controlled using the Benjamini-Hochberg procedure, applied within each analytic setting (aggregated, friendly, direct) across all 37 reported LIWC categories. The FDR-corrected summary tables (Tables \ref{tab:liwc_fdr_grammatical} and \ref{tab:liwc_fdr_semantic} below) report, for each category, Spearman’s $\rho$ between the participants’ and NAVI’s normalized frequencies, the uncorrected p-value, and the FDR-adjusted q-value, with asterisks denoting significance after the correction (* q < .05, ** q < .01, *** q < .001)}. Spearman's correlation ($\rho$) values are presented in the tables below for linguistic variables (Tables \ref{tab:lingVar.friendly.1} -- Table \ref{tab:lingVar.direct.2}) and semantic variables (Tables \ref{tab:semVar.friendly.1} -- Table \ref{tab:semVar.direct.3}) for interactions in both the friendly and direct condition. For clarity of presentation, p-values are reported using the standard asterisk notation (* p < 0.05, ** p < 0.01, *** p < 0.001). \dk{In the full correlation matrices below (Tables \ref{tab:lingVar.friendly.1} –\ref{tab:semVar.direct.3}), these asterisks correspond to uncorrected p-values; the matrices are retained for descriptive purposes, and significance after the FDR correction should be read from Tables \ref{tab:liwc_fdr_grammatical} and \ref{tab:liwc_fdr_semantic}} Additionally,  p\_ is used to represent participants' interactions and n\_ = \naviw's interactions in the following tables. The complete input and output datasets are available from the authors upon request. 

The LIWC categories, each represented by a code in tables \ref{tab:lingVar.friendly.1} -- \ref{tab:semVar.direct.3}, are described in detail in Table \ref{tab:liwc-vertical} below.

\begin{longtable}{@{}>{\raggedright\arraybackslash}p{.22\textwidth}
                    >{\raggedright\arraybackslash}p{.22\textwidth}
                    >{\raggedright\arraybackslash}p{.56\textwidth}@{}}
\caption{LIWC-style categories and descriptions.}\label{tab:liwc-vertical}\\
\toprule
\textbf{Code} & \textbf{Name} & \textbf{Description} \\
\midrule
\endfirsthead

\multicolumn{3}{@{}l}{\small\itshape Table \ref{tab:liwc-vertical} (continued)}\\
\toprule
\textbf{Code} & \textbf{Name} & \textbf{Description} \\
\midrule
\endhead

\bottomrule
\endfoot

\addlinespace
\multicolumn{3}{@{}l}{\textbf{General text metrics}}\\
\addlinespace
\texttt{WC} & Word Count & Number of words in the segment \\
\texttt{WPS} & Words per Sentence & Average words per sentence \\
\texttt{Sixltr} & Six or more letters & Percentage of words with six or more letters \\
\texttt{Dic} & Dictionary words & Percentage of words recognized in the LIWC dictionary \\

\addlinespace
\multicolumn{3}{@{}l}{\textbf{Grammar-related categories}}\\
\addlinespace
\texttt{funct} & Function words & Function words (articles, pronouns, prepositions, conjunctions, etc.) \\
\texttt{pronoun} & Pronouns & All pronouns \\
\texttt{ppron} & Personal pronouns & Personal pronouns \\
\texttt{i} & First person singular & I, me, my \\
\texttt{we} & First person plural & we, us, our \\
\texttt{you} & Second person & you, your \\
\texttt{shehe} & Third person singular & she, he, her, him \\
\texttt{they} & Third person plural & they, them, their \\
\texttt{ipron} & Impersonal pronouns & it, those, anything \\
\texttt{article} & Articles & a, an, the \\
\texttt{verb} & Verbs & All verbs \\
\texttt{auxverb} & Auxiliary verbs & be, have, will \\
\texttt{past} & Past tense & Past tense verbs \\
\texttt{present} & Present tense & Present tense verbs \\
\texttt{future} & Future tense & Future tense verbs \\
\texttt{adverb} & Adverbs & Adverbs \\
\texttt{preps} & Prepositions & Prepositions \\
\texttt{conj} & Conjunctions & Conjunctions \\
\texttt{negate} & Negations & no, not, never \\
\texttt{quant} & Quantifiers & many, few, much \\
\texttt{number} & Numbers & Numbers \\
\texttt{swear} & Swear words & Swear words \\

\addlinespace
\multicolumn{3}{@{}l}{\textbf{Semantic psychological categories}}\\
\addlinespace
\texttt{social} & Social processes & Social processes \\
\texttt{family} & Family & Family-related terms \\
\texttt{friend} & Friends & Friend-related terms \\
\texttt{humans} & Humans & Other human references \\
\texttt{affect} & Affect & Affective processes (emotions overall) \\
\texttt{posemo} & Positive emotion & Positive emotions \\
\texttt{negemo} & Negative emotion & Negative emotions \\
\texttt{anx} & Anxiety & Anxiety \\
\texttt{anger} & Anger & Anger \\
\texttt{sad} & Sadness & Sadness \\
\texttt{cogmech} & Cognitive processes & Cognitive processes \\
\texttt{insight} & Insight & Insight, self-reflection \\
\texttt{cause} & Causation & Causation \\
\texttt{discrep} & Discrepancy & Discrepancies, contradictions \\
\texttt{tentat} & Tentative & Tentative, uncertainty \\
\texttt{certain} & Certainty & Certainty \\
\texttt{inhib} & Inhibition & Inhibition, constraints \\
\texttt{incl} & Inclusion & Inclusion (with, include) \\
\texttt{excl} & Exclusion & Exclusion (but, without) \\
\texttt{percept} & Perceptual processes & Perceptual processes \\
\texttt{see} & See & Visual perception \\
\texttt{hear} & Hear & Auditory perception \\
\texttt{feel} & Feel & Sensory/feeling words \\
\texttt{bio} & Biological processes & Biological processes \\
\texttt{body} & Body & Body-related terms \\
\texttt{health} & Health & Health-related terms \\
\texttt{sexual} & Sexual & Sexual terms \\
\texttt{ingest} & Ingest & Eating/drinking \\
\texttt{relativ} & Relativity & Relativity (motion, space, time) \\
\texttt{motion} & Motion & Movement \\
\texttt{space} & Space & Spatial references \\
\texttt{time} & Time & Temporal references \\
\texttt{work} & Work & Work-related terms \\
\texttt{achieve} & Achievement & Achievement, success \\
\texttt{leisure} & Leisure & Leisure, hobbies \\
\texttt{home} & Home & Home, domestic life \\
\texttt{money} & Money & Money, finance \\
\texttt{relig} & Religion & Religion \\
\texttt{death} & Death & Death, mortality \\

\addlinespace
\multicolumn{3}{@{}l}{\textbf{Other categories}}\\
\addlinespace
\texttt{assent} & Assent/Agreement & yes, OK \\
\texttt{nonfl} & Nonfluencies & uh, um \\
\texttt{filler} & Fillers & you know, like \\

\addlinespace
\multicolumn{3}{@{}l}{\textbf{Punctuation}}\\
\addlinespace
\texttt{Period} & Periods & Periods (.) \\
\texttt{Comma} & Commas & Commas (,) \\
\texttt{Colon} & Colons & Colons (:) \\
\texttt{SemiC} & Semicolons & Semicolons (;) \\
\texttt{QMark} & Question marks & Question marks (?) \\
\texttt{Exclam} & Exclamation marks & Exclamation marks (!) \\
\texttt{Dash} & Dashes & Dashes (\textendash) \\
\texttt{Quote} & Quotation marks & Quotation marks (``\,'' ) \\
\texttt{Apostro} & Apostrophes & Apostrophes (\textquotesingle) \\
\texttt{Parenth} & Parentheses & Parentheses ( ) \\
\texttt{OtherP} & Other punctuation & Other punctuation \\
\texttt{AllPct} & All punctuation & All punctuation marks combined \\
\end{longtable}

\addtolength{\tabcolsep}{-3.5pt}
\begin{table}[htbp]
\centering
\caption{\dk{FDR-corrected Spearman correlations between participants' and NAVI's use of grammatical LIWC categories, for the aggregated dataset and separately for the friendly and direct conditions.}}
\label{tab:liwc_fdr_grammatical}
\small
\begin{tabular}{l rrr rrr rrr}
\toprule
 & \multicolumn{3}{c}{Aggregated} & \multicolumn{3}{c}{Friendly} & \multicolumn{3}{c}{Direct} \\
\cmidrule(lr){2-4} \cmidrule(lr){5-7} \cmidrule(lr){8-10}
LIWC category & $\rho$ & $p$ & $q$ & $\rho$ & $p$ & $q$ & $\rho$ & $p$ & $q$ \\
\midrule
\texttt{WPS} (Words per sentence) & .092 & .103 & .165 & $-$.064 & .426 & .606 & .088 & .270 & .356 \\
\texttt{funct} (Function words) & .005 & .930 & .956 & $-$.095 & .235 & .415 & .143 & .072 & .141 \\
\texttt{pronoun} (Pronouns) & .249*** & $<$.001 & $<$.001 & .119 & .138 & .283 & .252** & .001 & .006 \\
\texttt{ppron} (Personal pronouns) & .306*** & $<$.001 & $<$.001 & .127 & .111 & .256 & .457*** & $<$.001 & $<$.001 \\
\texttt{i} (First person singular) & .290*** & $<$.001 & $<$.001 & .043 & .593 & .707 & .296** & $<$.001 & .001 \\
\texttt{we} (First person plural) & .044 & .431 & .470 & .067 & .404 & .598 & .098 & .221 & .303 \\
\texttt{you} (Second person) & .019 & .738 & .780 & .052 & .513 & .654 & .204* & .010 & .031 \\
\texttt{verb} (Verbs) & .261*** & $<$.001 & $<$.001 & .152 & .056 & .160 & .271** & $<$.001 & .003 \\
\texttt{auxverb} (Auxiliary verbs) & .176** & .002 & .006 & $-$.049 & .539 & .665 & .281** & $<$.001 & .002 \\
\texttt{past} (Past tense) & .297*** & $<$.001 & $<$.001 & .241* & .002 & .017 & .136 & .088 & .156 \\
\texttt{present} (Present tense) & .144* & .010 & .028 & .124 & .121 & .264 & .171 & .032 & .074 \\
\texttt{future} (Future tense) & .044 & .432 & .470 & .074 & .354 & .546 & .038 & .637 & .694 \\
\texttt{conj} (Conjunctions) & $-$.228*** & $<$.001 & $<$.001 & $-$.059 & .465 & .614 & $-$.105 & .191 & .271 \\
\texttt{negate} (Negations) & .047 & .402 & .470 & .202 & .011 & .051 & $-$.002 & .975 & .975 \\
\texttt{quant} (Quantifiers) & .146* & .009 & .028 & .146 & .067 & .178 & .019 & .815 & .838 \\
\texttt{number} (Numbers) & .386*** & $<$.001 & $<$.001 & .357*** & $<$.001 & $<$.001 & .245** & .002 & .007 \\
\bottomrule
\end{tabular}

\medskip
\raggedright\footnotesize
\emph{Note.} Spearman rank-order correlations between participants' and NAVI's normalized frequencies of the same LIWC category (\texttt{p\_}$\times$\texttt{n\_} pairs). $p$ = uncorrected two-sided $p$-value; $q$ = Benjamini--Hochberg FDR-adjusted $p$-value, computed within each analytic setting (aggregated, friendly, direct) across all 37 reported LIWC categories (grammatical and semantic combined). Asterisks denote significance after FDR correction: * $q<.05$, ** $q<.01$, *** $q<.001$.
\end{table}
\addtolength{\tabcolsep}{3.5pt}

\addtolength{\tabcolsep}{-3.5pt}
\begin{table}[htbp]
\centering
\caption{\dk{FDR-corrected Spearman correlations between participants' and NAVI's use of semantic LIWC categories, for the aggregated dataset and separately for the friendly and direct conditions.}}
\label{tab:liwc_fdr_semantic}
\small
\begin{tabular}{l rrr rrr rrr}
\toprule
 & \multicolumn{3}{c}{Aggregated} & \multicolumn{3}{c}{Friendly} & \multicolumn{3}{c}{Direct} \\
\cmidrule(lr){2-4} \cmidrule(lr){5-7} \cmidrule(lr){8-10}
LIWC category & $\rho$ & $p$ & $q$ & $\rho$ & $p$ & $q$ & $\rho$ & $p$ & $q$ \\
\midrule
\texttt{social} (Social processes) & .127 & .024 & .055 & .162 & .042 & .157 & .251** & .001 & .006 \\
\texttt{affect} (Affect) & .145* & .010 & .028 & .237* & .003 & .017 & .415*** & $<$.001 & $<$.001 \\
\texttt{posemo} (Positive emotion) & .142* & .012 & .029 & .263** & $<$.001 & .009 & .406*** & $<$.001 & $<$.001 \\
\texttt{negemo} (Negative emotion) & .156* & .006 & .019 & .231* & .004 & .019 & .079 & .323 & .398 \\
\texttt{cogmech} (Cognitive processes) & .081 & .150 & .214 & $-$.014 & .864 & .913 & .136 & .088 & .156 \\
\texttt{insight} (Insight) & .123 & .029 & .059 & .111 & .164 & .320 & .060 & .454 & .509 \\
\texttt{cause} (Causation) & $-$.070 & .214 & .282 & $-$.030 & .705 & .791 & $-$.121 & .129 & .199 \\
\texttt{discrep} (Discrepancy) & .088 & .120 & .177 & .085 & .287 & .462 & $-$.033 & .681 & .720 \\
\texttt{tentat} (Tentative) & $-$.045 & .424 & .470 & .035 & .664 & .768 & .151 & .058 & .126 \\
\texttt{certain} (Certainty) & .126 & .026 & .056 & $-$.001 & .986 & .986 & .184 & .020 & .054 \\
\texttt{inhib} (Inhibition) & .068 & .225 & .288 & .103 & .197 & .365 & .171 & .031 & .074 \\
\texttt{incl} (Inclusion) & .118 & .036 & .070 & .059 & .462 & .614 & .197* & .013 & .037 \\
\texttt{excl} (Exclusion) & $-$.055 & .327 & .404 & $-$.162 & .042 & .157 & .108 & .177 & .262 \\
\texttt{percept} (Perceptual processes) & .079 & .160 & .220 & .152 & .056 & .160 & .080 & .317 & .398 \\
\texttt{see} (See) & .105 & .062 & .109 & .259** & $<$.001 & .009 & .074 & .358 & .427 \\
\texttt{relativ} (Relativity) & .211*** & $<$.001 & $<$.001 & .090 & .259 & .436 & .212* & .008 & .025 \\
\texttt{motion} (Motion) & .110 & .050 & .092 & $-$.014 & .862 & .913 & .124 & .120 & .194 \\
\texttt{space} (Space) & .090 & .112 & .172 & .276** & $<$.001 & .008 & .147 & .065 & .133 \\
\texttt{time} (Time) & $-$.096 & .088 & .149 & .005 & .950 & .977 & .062 & .442 & .509 \\
\texttt{work} (Work) & .425*** & $<$.001 & $<$.001 & .157 & .049 & .160 & .564*** & $<$.001 & $<$.001 \\
\texttt{achieve} (Achievement) & .000 & .996 & .996 & .135 & .090 & .223 & $-$.127 & .112 & .189 \\
\bottomrule
\end{tabular}

\medskip
\raggedright\footnotesize
\emph{Note.} Spearman rank-order correlations between participants' and NAVI's normalized frequencies of the same LIWC category (\texttt{p\_}$\times$\texttt{n\_} pairs). $p$ = uncorrected two-sided $p$-value; $q$ = Benjamini--Hochberg FDR-adjusted $p$-value, computed within each analytic setting (aggregated, friendly, direct) across all 37 reported LIWC categories (grammatical and semantic combined). Asterisks denote significance after FDR correction: * $q<.05$, ** $q<.01$, *** $q<.001$.
\end{table}
\addtolength{\tabcolsep}{3.5pt}

\addtolength{\tabcolsep}{-3.5pt}
\begin{table}[p]
  \centering
  \caption{Spearman's correlation of linguistic variables in interactions in the friendly condition (Part 1).}
  \label{tab:lingVar.friendly.1}
  \scriptsize
  \begin{tabular}{lcccccccc}
    \toprule
    Variable & n\_WPS & n\_funct & n\_pronoun & n\_ppron & n\_i & n\_we & n\_you & n\_verb \\
    \midrule
    17.\ p\_WPS      & -0.064 & 0.160\textsuperscript{*} & 0.186\textsuperscript{*} & 0.125 & 0.122 & 0.083 & 0.095 & -0.029 \\
    18.\ p\_funct    & -0.065 & -0.095 & -0.067 & -0.122 & -0.014 & -0.025 & -0.185\textsuperscript{*} & -0.038 \\
    19.\ p\_pronoun  & -0.179\textsuperscript{*} & 0.029 & 0.119 & 0.127 & 0.181\textsuperscript{*} & 0.006 & 0.041 & 0.161\textsuperscript{*} \\
    20.\ p\_ppron    & -0.204\textsuperscript{*} & -0.032 & 0.106 & 0.127 & 0.099 & -0.025 & 0.130 & 0.211\textsuperscript{**} \\
    21.\ p\_i        & -0.097 & -0.028 & 0.045 & 0.068 & 0.043 & -0.058 & 0.077 & 0.209\textsuperscript{**} \\
    22.\ p\_we       & -0.094 & 0.086 & 0.052 & 0.126 & 0.088 & 0.067 & 0.128 & 0.151 \\
    23.\ p\_you      & -0.189\textsuperscript{*} & -0.064 & 0.127 & 0.065 & 0.059 & 0.078 & 0.052 & -0.036 \\
    24.\ p\_verb     & -0.156 & -0.054 & 0.102 & 0.125 & 0.174\textsuperscript{*} & -0.083 & 0.081 & 0.152 \\
    25.\ p\_auxverb  & -0.125 & -0.143 & -0.008 & -0.088 & 0.074 & -0.026 & -0.150 & 0.035 \\
    26.\ p\_past     & -0.0006 & 0.124 & 0.082 & 0.152 & 0.126 & 0.030 & 0.129 & -0.047 \\
    27.\ p\_present  & -0.156 & -0.095 & 0.074 & 0.024 & 0.110 & -0.086 & -0.014 & 0.161\textsuperscript{*} \\
    28.\ p\_future   & 0.022 & -0.123 & 0.010 & 0.052 & 0.030 & 0.018 & 0.033 & -0.012 \\
    29.\ p\_conj     & 0.168\textsuperscript{*} & 0.010 & 0.084 & 0.108 & 0.039 & 0.025 & 0.093 & 0.040 \\
    30.\ p\_negate   & 0.066 & 0.018 & 0.050 & -0.044 & 0.019 & -0.058 & -0.068 & -0.045 \\
    31.\ p\_quant    & 0.333\textsuperscript{***} & -0.108 & -0.030 & -0.107 & -0.112 & 0.057 & -0.108 & -0.042 \\
    32.\ p\_number   & -0.034 & 0.071 & -0.148 & -0.207\textsuperscript{**} & -0.116 & -0.097 & -0.213\textsuperscript{**} & -0.326\textsuperscript{***} \\
    \bottomrule
  \end{tabular}
\end{table}
\addtolength{\tabcolsep}{3.5pt}

\addtolength{\tabcolsep}{-3.5pt}
\begin{table}[p]
  \centering
  \caption{Spearman's correlation of linguistic variables in interactions in the friendly condition (Part 2).}
  \label{tab:lingVar.friendly.2}
  \scriptsize
  \begin{tabular}{lcccccccc}
    \toprule
    Variable & n\_auxverb & n\_past & n\_present & n\_future & n\_conj & n\_negate & n\_quant & n\_number \\
    \midrule
    17.\ p\_WPS      & 0.008 & -0.097 & 0.044 & -0.045 & 0.051 & -0.151 & -0.088 & 0.036 \\
    18.\ p\_funct    & 0.004 & -0.045 & -0.029 & 0.034 & -0.094 & 0.157\textsuperscript{*} & -0.084 & 0.057 \\
    19.\ p\_pronoun  & 0.037 & 0.190\textsuperscript{*} & 0.075 & 0.056 & 0.073 & 0.101 & 0.055 & 0.159\textsuperscript{*} \\
    20.\ p\_ppron    & -0.031 & 0.116 & 0.189\textsuperscript{*} & -0.031 & 0.098 & -0.002 & 0.077 & -0.001 \\
    21.\ p\_i        & -0.016 & 0.133 & 0.174\textsuperscript{*} & -0.008 & 0.125 & 0.013 & 0.065 & 0.005 \\
    22.\ p\_we       & 0.057 & 0.163\textsuperscript{*} & 0.090 & -0.015 & -0.020 & -0.136 & -0.094 & -0.016 \\
    23.\ p\_you      & -0.092 & -0.022 & 0.015 & -0.105 & 0.045 & -0.010 & 0.061 & -0.019 \\
    24.\ p\_verb     & -0.012 & 0.183\textsuperscript{*} & 0.082 & 0.047 & 0.073 & 0.052 & 0.138 & -0.059 \\
    25.\ p\_auxverb  & -0.049 & -0.039 & -0.048 & 0.120 & 0.104 & 0.150 & 0.068 & -0.097 \\
    26.\ p\_past     & 0.137 & 0.241\textsuperscript{**} & -0.097 & 0.031 & -0.063 & 0.005 & 0.088 & 0.123 \\
    27.\ p\_present  & -0.025 & 0.027 & 0.124 & 0.011 & 0.099 & 0.076 & 0.115 & -0.125 \\
    28.\ p\_future   & -0.029 & 0.117 & -0.076 & 0.074 & -0.023 & -0.071 & 0.053 & -0.123 \\
    29.\ p\_conj     & 0.138 & -0.067 & 0.054 & 0.121 & -0.059 & 0.083 & -0.122 & 0.026 \\
    30.\ p\_negate   & 0.131 & -0.011 & -0.004 & -0.093 & -0.104 & 0.202\textsuperscript{*} & 0.078 & -0.006 \\
    31.\ p\_quant    & 0.062 & 0.098 & -0.077 & -0.020 & 0.055 & -0.021 & 0.146 & -0.145 \\
    32.\ p\_number   & -0.127 & -0.036 & -0.252\textsuperscript{**} & -0.065 & -0.257\textsuperscript{**} & 0.029 & -0.168\textsuperscript{*} & 0.357\textsuperscript{***} \\
    \bottomrule
  \end{tabular}
\end{table}
\addtolength{\tabcolsep}{3.5pt}

\addtolength{\tabcolsep}{-3.5pt}
\begin{table}[p]
  \centering
  \caption{Spearman's correlation of linguistic variables in interactions in the direct condition (Part 1).}
  \label{tab:lingVar.direct.1}
  \scriptsize
  \begin{tabular}{lcccccccc}
    \toprule
    Variable & n\_WPS & n\_funct & n\_pronoun & n\_ppron & n\_i & n\_we & n\_you & n\_verb \\
    \midrule
    17.\ p\_WPS      & 0.088 & 0.066 & -0.117 & 0.032 & 0.028 & 0.155 & 0.010 & -0.050 \\
    18.\ p\_funct    & 0.218\textsuperscript{**} & 0.143 & -0.002 & -0.144 & 0.004 & -0.120 & -0.147 & -0.047 \\
    19.\ p\_pronoun  & -0.258\textsuperscript{**} & -0.087 & 0.252\textsuperscript{**} & 0.339\textsuperscript{***} & 0.236\textsuperscript{**} & 0.029 & 0.343\textsuperscript{***} & 0.287\textsuperscript{***} \\
    20.\ p\_ppron    & -0.419\textsuperscript{***} & -0.250\textsuperscript{**} & 0.171\textsuperscript{*} & 0.457\textsuperscript{***} & 0.252\textsuperscript{**} & 0.073 & 0.453\textsuperscript{***} & 0.319\textsuperscript{***} \\
    21.\ p\_i        & -0.357\textsuperscript{***} & -0.186\textsuperscript{*} & 0.117 & 0.421\textsuperscript{***} & 0.296\textsuperscript{***} & 0.051 & 0.384\textsuperscript{***} & 0.328\textsuperscript{***} \\
    22.\ p\_we       & 0.012 & -0.011 & -0.148 & -0.115 & -0.007 & 0.098 & -0.160\textsuperscript{*} & -0.063 \\
    23.\ p\_you      & -0.163\textsuperscript{*} & -0.111 & 0.173\textsuperscript{*} & 0.134 & -0.053 & 0.093 & 0.204\textsuperscript{**} & 0.049 \\
    24.\ p\_verb     & -0.237\textsuperscript{**} & 0.038 & 0.248\textsuperscript{**} & 0.238\textsuperscript{**} & 0.139 & -0.032 & 0.258\textsuperscript{**} & 0.271\textsuperscript{***} \\
    25.\ p\_auxverb  & 0.071 & 0.267\textsuperscript{***} & 0.260\textsuperscript{***} & 0.073 & 0.095 & -0.064 & 0.083 & 0.161\textsuperscript{*} \\
    26.\ p\_past     & -0.173\textsuperscript{*} & -0.030 & 0.059 & 0.096 & -0.079 & 0.022 & 0.178\textsuperscript{*} & 0.134 \\
    27.\ p\_present  & -0.158\textsuperscript{*} & 0.044 & 0.226\textsuperscript{**} & 0.170\textsuperscript{*} & 0.151 & -0.033 & 0.166\textsuperscript{*} & 0.247\textsuperscript{**} \\
    28.\ p\_future   & -0.048 & -0.049 & -0.006 & 0.048 & -0.050 & 0.062 & 0.065 & -0.018 \\
    29.\ p\_conj     & 0.085 & -0.075 & -0.156 & -0.172\textsuperscript{*} & -0.208\textsuperscript{**} & -0.028 & -0.146 & -0.077 \\
    30.\ p\_negate   & 0.083 & -0.108 & -0.087 & -0.068 & -0.119 & 0.131 & -0.048 & -0.139 \\
    31.\ p\_quant    & 0.169\textsuperscript{*} & -0.007 & -0.107 & -0.203\textsuperscript{*} & -0.235\textsuperscript{**} & 0.136 & -0.193\textsuperscript{*} & -0.207\textsuperscript{**} \\
    32.\ p\_number   & 0.215\textsuperscript{**} & 0.205\textsuperscript{**} & -0.158\textsuperscript{*} & -0.254\textsuperscript{**} & -0.130 & -0.015 & -0.244\textsuperscript{**} & -0.252\textsuperscript{**} \\
    \bottomrule
  \end{tabular}
\end{table}
\addtolength{\tabcolsep}{3.5pt}

\addtolength{\tabcolsep}{-3.5pt}
\begin{table}[p]
  \centering
  \caption{Spearman's correlation of linguistic variables in interactions in the direct condition (Part 2).}
  \label{tab:lingVar.direct.2}
  \scriptsize
  \begin{tabular}{lcccccccc}
    \toprule
    Variable & n\_auxverb & n\_past & n\_present & n\_future & n\_conj & n\_negate & n\_quant & n\_number \\
    \midrule
    17.\ p\_WPS      & -0.089 & 0.026 & -0.065 & -0.053 & 0.037 & -0.094 & 0.137 & 0.199\textsuperscript{*} \\
    18.\ p\_funct    & 0.028 & -0.055 & -0.084 & 0.102 & -0.082 & 0.033 & -0.115 & 0.036 \\
    19.\ p\_pronoun  & 0.120 & 0.009 & 0.207\textsuperscript{**} & 0.149 & 0.129 & -0.147 & 0.046 & -0.103 \\
    20.\ p\_ppron    & 0.077 & -0.076 & 0.220\textsuperscript{**} & 0.187\textsuperscript{*} & 0.258\textsuperscript{**} & -0.297\textsuperscript{***} & 0.015 & -0.024 \\
    21.\ p\_i        & 0.070 & -0.007 & 0.208\textsuperscript{**} & 0.213\textsuperscript{**} & 0.303\textsuperscript{***} & -0.290\textsuperscript{***} & 0.012 & -0.044 \\
    22.\ p\_we       & -0.152 & -0.050 & -0.015 & -0.073 & -0.173\textsuperscript{*} & -0.002 & -0.083 & 0.114 \\
    23.\ p\_you      & 0.044 & -0.067 & 0.054 & -0.0001 & -0.091 & 0.009 & 0.050 & -0.017 \\
    24.\ p\_verb     & 0.332\textsuperscript{***} & 0.138 & 0.187\textsuperscript{*} & 0.201\textsuperscript{*} & 0.066 & 0.010 & 0.140 & -0.157\textsuperscript{*} \\
    25.\ p\_auxverb  & 0.281\textsuperscript{***} & 0.067 & 0.051 & 0.142 & -0.023 & 0.184\textsuperscript{*} & 0.153 & -0.045 \\
    26.\ p\_past     & 0.110 & 0.136 & 0.137 & 0.020 & 0.014 & 0.063 & 0.017 & -0.072 \\
    27.\ p\_present  & 0.272\textsuperscript{***} & 0.125 & 0.171\textsuperscript{*} & 0.173\textsuperscript{*} & 0.047 & -0.016 & 0.118 & -0.168\textsuperscript{*} \\
    28.\ p\_future   & 0.120 & -0.114 & -0.023 & 0.038 & -0.052 & 0.155 & 0.103 & -0.034 \\
    29.\ p\_conj     & -0.104 & 0.080 & -0.075 & 0.035 & -0.105 & -0.039 & 0.028 & 0.067 \\
    30.\ p\_negate   & -0.177\textsuperscript{*} & -0.004 & -0.153 & 0.063 & -0.078 & -0.002 & 0.116 & 0.195\textsuperscript{*} \\
    31.\ p\_quant    & -0.189\textsuperscript{*} & 0.020 & -0.124 & -0.240\textsuperscript{**} & -0.128 & 0.227\textsuperscript{**} & 0.019 & 0.139 \\
    32.\ p\_number   & -0.134 & -0.058 & -0.133 & -0.275\textsuperscript{***} & -0.222\textsuperscript{**} & 0.186\textsuperscript{*} & -0.113 & 0.245\textsuperscript{**} \\
    \bottomrule
  \end{tabular}
\end{table}
\addtolength{\tabcolsep}{3.5pt}

\addtolength{\tabcolsep}{-3.5pt}
\begin{table}[p]
  \centering
  \caption{Spearman's correlation of semantic variables in interactions in the friendly condition (Part 1).}
  \label{tab:semVar.friendly.1}
  \scriptsize
  \begin{tabular}{lccccccc}
    \toprule
    Variable & n\_social & n\_affect & n\_posemo & n\_negemo & n\_cogmech & n\_insight & n\_cause \\
    \midrule
    22.\ p\_social   & 0.162\textsuperscript{*} & 0.210\textsuperscript{**} & 0.186\textsuperscript{*} & 0.067 & 0.075 & -0.037 & 0.048 \\
    23.\ p\_affect   & 0.226\textsuperscript{**} & 0.237\textsuperscript{**} & 0.247\textsuperscript{**} & 0.020 & 0.111 & 0.060 & -0.126 \\
    24.\ p\_posemo   & 0.218\textsuperscript{**} & 0.230\textsuperscript{**} & 0.263\textsuperscript{***} & -0.027 & 0.123 & 0.063 & -0.112 \\
    25.\ p\_negemo   & -0.026 & 0.076 & -0.032 & 0.231\textsuperscript{**} & -0.075 & 0.021 & -0.127 \\
    26.\ p\_cogmech  & 0.013 & -0.028 & -0.093 & 0.137 & -0.014 & -0.044 & -0.121 \\
    27.\ p\_insight  & 0.019 & 0.093 & 0.103 & -0.052 & 0.047 & 0.111 & -0.111 \\
    28.\ p\_cause    & 0.023 & -0.170\textsuperscript{*} & -0.210\textsuperscript{**} & 0.058 & 0.021 & 0.061 & -0.030 \\
    29.\ p\_discrep  & 0.119 & 0.154 & 0.124 & 0.018 & 0.105 & -0.008 & -0.0009 \\
    30.\ p\_tentat   & -0.121 & -0.078 & -0.069 & -0.018 & -0.026 & 0.175\textsuperscript{*} & -0.152 \\
    31.\ p\_certain  & -0.0007 & 0.055 & 0.034 & 0.093 & -0.054 & 0.149 & -0.133 \\
    32.\ p\_inhib    & -0.035 & -0.077 & -0.105 & 0.066 & -0.072 & -0.083 & -0.020 \\
    33.\ p\_incl     & -0.098 & -0.053 & -0.091 & 0.114 & -0.140 & -0.141 & -0.177\textsuperscript{*} \\
    34.\ p\_excl     & -0.056 & -0.187\textsuperscript{*} & -0.268\textsuperscript{***} & 0.159\textsuperscript{*} & -0.201\textsuperscript{*} & 0.018 & -0.117 \\
    35.\ p\_percept  & -0.046 & -0.031 & -0.021 & -0.078 & 0.074 & 0.091 & -0.128 \\
    36.\ p\_see      & -0.046 & -0.030 & -0.023 & -0.073 & 0.081 & 0.080 & -0.128 \\
    37.\ p\_relativ  & -0.079 & -0.130 & -0.129 & -0.067 & -0.172\textsuperscript{*} & 0.006 & -0.085 \\
    38.\ p\_motion   & 0.0007 & -0.151 & -0.138 & -0.070 & 0.030 & 0.136 & -0.125 \\
    39.\ p\_space    & -0.164\textsuperscript{*} & -0.081 & -0.076 & -0.055 & -0.175\textsuperscript{*} & 0.030 & -0.091 \\
    40.\ p\_time     & -0.034 & -0.048 & -0.046 & -0.087 & 0.007 & 0.075 & -0.045 \\
    41.\ p\_work     & -0.060 & -0.194\textsuperscript{*} & -0.197\textsuperscript{*} & 0.022 & -0.117 & -0.054 & -0.133 \\
    42.\ p\_achieve  & -0.087 & 0.089 & 0.071 & 0.013 & -0.109 & 0.105 & -0.054 \\
    \bottomrule
  \end{tabular}
\end{table}
\addtolength{\tabcolsep}{3.5pt}

\addtolength{\tabcolsep}{-3.5pt}
\begin{table}[p]
  \centering
  \caption{Spearman's correlation of semantic variables in interactions in the friendly condition (Part 2).}
  \label{tab:semVar.friendly.2}
  \scriptsize
  \begin{tabular}{lccccccc}
    \toprule
    Variable & n\_discrep & n\_tentat & n\_certain & n\_inhib & n\_incl & n\_excl & n\_percept \\
    \midrule
    22.\ p\_social   & -0.002 & 0.023 & -0.007 & 0.145 & 0.131 & -0.024 & -0.019 \\
    23.\ p\_affect   & 0.015 & 0.085 & 0.060 & 0.042 & 0.139 & 0.063 & -0.063 \\
    24.\ p\_posemo   & 0.040 & 0.065 & 0.071 & 0.040 & 0.142 & 0.055 & -0.063 \\
    25.\ p\_negemo   & -0.159\textsuperscript{*} & 0.065 & -0.045 & -0.042 & -0.004 & -0.027 & 0.012 \\
    26.\ p\_cogmech  & -0.149 & -0.018 & -0.044 & 0.178\textsuperscript{*} & 0.062 & -0.056 & 0.010 \\
    27.\ p\_insight  & -0.121 & 0.0010 & 0.034 & 0.105 & 0.116 & -0.031 & 0.031 \\
    28.\ p\_cause    & -0.003 & 0.033 & -0.187\textsuperscript{*} & 0.324\textsuperscript{***} & -0.022 & 0.016 & 0.032 \\
    29.\ p\_discrep  & 0.085 & 0.128 & 0.027 & -0.0004 & -0.027 & 0.113 & -0.119 \\
    30.\ p\_tentat   & -0.058 & 0.035 & 0.038 & -0.079 & -0.149 & -0.060 & 0.094 \\
    31.\ p\_certain  & -0.170\textsuperscript{*} & -0.062 & -0.001 & 0.050 & 0.024 & -0.171\textsuperscript{*} & 0.034 \\
    32.\ p\_inhib    & -0.104 & -0.108 & 0.053 & 0.103 & 0.036 & -0.045 & -0.046 \\
    33.\ p\_incl     & -0.117 & -0.125 & 0.016 & -0.125 & 0.059 & -0.064 & 0.044 \\
    34.\ p\_excl     & -0.238\textsuperscript{**} & -0.085 & -0.023 & -0.082 & -0.132 & -0.162\textsuperscript{*} & 0.085 \\
    35.\ p\_percept  & 0.104 & 0.060 & 0.060 & 0.011 & 0.041 & 0.045 & 0.152 \\
    36.\ p\_see      & 0.107 & 0.058 & 0.077 & 0.013 & 0.038 & 0.053 & 0.190\textsuperscript{*} \\
    37.\ p\_relativ  & 0.016 & -0.109 & -0.050 & -0.310\textsuperscript{***} & -0.153 & -0.005 & -0.041 \\
    38.\ p\_motion   & 0.037 & -0.009 & -0.058 & -0.069 & -0.017 & 0.068 & 0.070 \\
    39.\ p\_space    & -0.016 & -0.130 & 0.012 & -0.333\textsuperscript{***} & -0.126 & -0.048 & 0.009 \\
    40.\ p\_time     & 0.065 & -0.042 & -0.017 & -0.086 & -0.090 & 0.029 & -0.128 \\
    41.\ p\_work     & -0.035 & 0.018 & -0.078 & -0.140 & -0.071 & -0.060 & 0.066 \\
    42.\ p\_achieve  & -0.267\textsuperscript{***} & -0.132 & 0.098 & -0.049 & -0.046 & -0.124 & -0.053 \\
    \bottomrule
  \end{tabular}
\end{table}
\addtolength{\tabcolsep}{3.5pt}

\addtolength{\tabcolsep}{-3.5pt}
\begin{table}[p]
  \centering
  \caption{Spearman's correlation of semantic variables in interactions in the friendly condition (Part 3).}
  \label{tab:semVar.friendly.3}
  \scriptsize
  \begin{tabular}{lccccccc}
    \toprule
    Variable & n\_see & n\_relativ & n\_motion & n\_space & n\_time & n\_work & n\_achieve \\
    \midrule
    22.\ p\_social   & -0.085 & -0.107 & 0.108 & -0.142 & -0.118 & -0.090 & 0.024 \\
    23.\ p\_affect   & -0.043 & -0.133 & 0.130 & -0.285\textsuperscript{***} & 0.053 & -0.111 & 0.081 \\
    24.\ p\_posemo   & -0.037 & -0.145 & 0.096 & -0.282\textsuperscript{***} & 0.044 & -0.080 & 0.070 \\
    25.\ p\_negemo   & -0.035 & 0.046 & 0.086 & 0.050 & -0.010 & -0.108 & 0.061 \\
    26.\ p\_cogmech  & -0.002 & 0.006 & 0.178\textsuperscript{*} & -0.017 & -0.055 & -0.044 & -0.062 \\
    27.\ p\_insight  & 0.040 & -0.088 & 0.030 & -0.087 & 0.058 & -0.010 & -0.022 \\
    28.\ p\_cause    & 0.009 & 0.105 & 0.306\textsuperscript{***} & 0.025 & -0.093 & -0.127 & -0.009 \\
    29.\ p\_discrep  & -0.132 & -0.225\textsuperscript{**} & 0.005 & -0.155 & -0.055 & -0.044 & -0.062 \\
    30.\ p\_tentat   & 0.109 & -0.035 & -0.007 & 0.040 & -0.073 & 0.079 & -0.018 \\
    31.\ p\_certain  & -0.022 & 0.084 & 0.088 & 0.025 & -0.011 & -0.021 & 0.056 \\
    32.\ p\_inhib    & -0.021 & 0.186\textsuperscript{*} & 0.089 & 0.197\textsuperscript{*} & -0.070 & 0.045 & 0.091 \\
    33.\ p\_incl     & 0.036 & 0.071 & -0.047 & 0.144 & -0.120 & 0.097 & -0.071 \\
    34.\ p\_excl     & 0.113 & -0.039 & 0.076 & 0.015 & -0.175\textsuperscript{*} & -0.014 & 0.035 \\
    35.\ p\_percept  & 0.223\textsuperscript{**} & -0.113 & -0.150 & -0.007 & -0.148 & 0.183\textsuperscript{*} & 0.126 \\
    36.\ p\_see      & 0.259\textsuperscript{***} & -0.117 & -0.130 & -0.015 & -0.147 & 0.180\textsuperscript{*} & 0.155 \\
    37.\ p\_relativ  & -0.092 & 0.090 & -0.064 & 0.177\textsuperscript{*} & -0.069 & -0.003 & -0.151 \\
    38.\ p\_motion   & -0.024 & -0.005 & -0.014 & 0.002 & -0.026 & -0.036 & -0.133 \\
    39.\ p\_space    & -0.039 & 0.102 & -0.119 & 0.276\textsuperscript{***} & -0.167\textsuperscript{*} & 0.101 & -0.064 \\
    40.\ p\_time     & -0.095 & -0.092 & 0.003 & -0.080 & 0.005 & -0.071 & -0.173\textsuperscript{*} \\
    41.\ p\_work     & 0.070 & 0.020 & 0.010 & 0.054 & -0.085 & 0.157\textsuperscript{*} & -0.122 \\
    42.\ p\_achieve  & -0.004 & -0.004 & -0.045 & 0.027 & 0.026 & 0.014 & 0.135 \\
    \bottomrule
  \end{tabular}
\end{table}
\addtolength{\tabcolsep}{3.5pt}

\addtolength{\tabcolsep}{-3.5pt}
\begin{table}[p]
  \centering
  \caption{Spearman's correlation of semantic variables in interactions in the direct condition (Part 1).}
  \label{tab:semVar.direct.1}
  \scriptsize
  \begin{tabular}{lccccccc}
    \toprule
    Variable & n\_social & n\_affect & n\_posemo & n\_negemo & n\_cogmech & n\_insight & n\_cause \\
    \midrule
    22.\ p\_social   & 0.251\textsuperscript{**} & 0.447\textsuperscript{***} & 0.461\textsuperscript{***} & 0.089 & 0.123 & 0.135 & -0.060 \\
    23.\ p\_affect   & 0.368\textsuperscript{***} & 0.415\textsuperscript{***} & 0.421\textsuperscript{***} & 0.108 & 0.038 & 0.048 & 0.022 \\
    24.\ p\_posemo   & 0.364\textsuperscript{***} & 0.397\textsuperscript{***} & 0.406\textsuperscript{***} & 0.103 & 0.040 & 0.036 & 0.042 \\
    25.\ p\_negemo   & -0.093 & 0.118 & 0.079 & 0.079 & 0.041 & 0.070 & -0.185\textsuperscript{*} \\
    26.\ p\_cogmech  & 0.021 & 0.041 & 0.052 & -0.001 & 0.136 & -0.023 & -0.068 \\
    27.\ p\_insight  & 0.059 & 0.049 & 0.042 & 0.008 & 0.077 & 0.060 & 0.025 \\
    28.\ p\_cause    & -0.096 & -0.068 & -0.096 & 0.086 & 0.023 & 0.005 & -0.121 \\
    29.\ p\_discrep  & -0.003 & 0.047 & 0.024 & 0.034 & -0.033 & -0.043 & 0.007 \\
    30.\ p\_tentat   & -0.254\textsuperscript{**} & -0.131 & -0.126 & -0.052 & 0.031 & -0.009 & -0.083 \\
    31.\ p\_certain  & -0.050 & -0.014 & 0.025 & -0.051 & 0.123 & 0.027 & -0.249\textsuperscript{**} \\
    32.\ p\_inhib    & -0.014 & 0.022 & 0.036 & -0.042 & 0.194\textsuperscript{*} & -0.117 & -0.208\textsuperscript{**} \\
    33.\ p\_incl     & -0.074 & 0.158\textsuperscript{*} & 0.162\textsuperscript{*} & 0.040 & 0.034 & -0.009 & -0.200\textsuperscript{*} \\
    34.\ p\_excl     & -0.194\textsuperscript{*} & -0.088 & -0.077 & -0.069 & 0.084 & -0.018 & -0.154 \\
    35.\ p\_percept  & -0.054 & 0.239\textsuperscript{**} & 0.216\textsuperscript{**} & 0.024 & 0.063 & 0.012 & -0.072 \\
    36.\ p\_see      & -0.042 & 0.260\textsuperscript{***} & 0.234\textsuperscript{**} & 0.036 & 0.065 & 0.025 & -0.074 \\
    37.\ p\_relativ  & -0.114 & -0.219\textsuperscript{**} & -0.214\textsuperscript{**} & -0.050 & -0.285\textsuperscript{***} & -0.237\textsuperscript{**} & 0.065 \\
    38.\ p\_motion   & -0.077 & -0.096 & -0.091 & -0.025 & -0.133 & -0.151 & -0.017 \\
    39.\ p\_space    & -0.274\textsuperscript{***} & -0.214\textsuperscript{**} & -0.194\textsuperscript{*} & -0.107 & -0.014 & -0.058 & -0.169\textsuperscript{*} \\
    40.\ p\_time     & 0.184\textsuperscript{*} & 0.048 & 0.036 & 0.002 & -0.275\textsuperscript{***} & -0.213\textsuperscript{**} & 0.213\textsuperscript{**} \\
    41.\ p\_work     & -0.381\textsuperscript{***} & -0.059 & -0.109 & 0.112 & 0.015 & 0.037 & -0.267\textsuperscript{***} \\
    42.\ p\_achieve  & -0.233\textsuperscript{**} & -0.0007 & -0.084 & 0.206\textsuperscript{**} & 0.046 & 0.107 & -0.117 \\
    \bottomrule
  \end{tabular}
\end{table}

\addtolength{\tabcolsep}{-3.5pt}
\begin{table}[p]
  \centering
  \caption{Spearman's correlation of semantic variables in interactions in the direct condition (Part 2).}
  \label{tab:semVar.direct.2}
  \scriptsize
  \begin{tabular}{lccccccc}
    \toprule
    Variable & n\_discrep & n\_tentat & n\_certain & n\_inhib & n\_incl & n\_excl & n\_percept \\
    \midrule
    22.\ p\_social   & 0.058 & 0.099 & 0.088 & 0.202\textsuperscript{*} & -0.164\textsuperscript{*} & 0.041 & -0.055 \\
    23.\ p\_affect   & 0.063 & 0.176\textsuperscript{*} & -0.047 & 0.251\textsuperscript{**} & -0.250\textsuperscript{**} & 0.045 & 0.121 \\
    24.\ p\_posemo   & 0.063 & 0.148 & -0.050 & 0.261\textsuperscript{***} & -0.255\textsuperscript{**} & 0.042 & 0.112 \\
    25.\ p\_negemo   & 0.040 & 0.210\textsuperscript{**} & 0.127 & -0.090 & 0.109 & 0.064 & 0.045 \\
    26.\ p\_cogmech  & 0.093 & 0.037 & 0.094 & 0.124 & 0.058 & -0.022 & -0.107 \\
    27.\ p\_insight  & 0.083 & 0.092 & 0.081 & 0.038 & -0.0004 & -0.043 & -0.080 \\
    28.\ p\_cause    & 0.076 & -0.009 & 0.037 & 0.043 & -0.044 & -0.019 & -0.071 \\
    29.\ p\_discrep  & -0.033 & -0.005 & 0.033 & -0.048 & 0.038 & -0.086 & -0.051 \\
    30.\ p\_tentat   & -0.015 & 0.151 & 0.074 & -0.132 & 0.088 & 0.077 & -0.096 \\
    31.\ p\_certain  & 0.247\textsuperscript{**} & 0.004 & 0.184\textsuperscript{*} & -0.007 & 0.032 & 0.053 & -0.205\textsuperscript{**} \\
    32.\ p\_inhib    & 0.063 & 0.179\textsuperscript{*} & 0.083 & 0.171\textsuperscript{*} & -0.025 & 0.173\textsuperscript{*} & -0.158\textsuperscript{*} \\
    33.\ p\_incl     & -0.092 & 0.008 & 0.139 & -0.062 & 0.197\textsuperscript{*} & 0.028 & -0.046 \\
    34.\ p\_excl     & 0.044 & 0.091 & 0.159\textsuperscript{*} & -0.070 & 0.013 & 0.108 & -0.137 \\
    35.\ p\_percept  & -0.018 & -0.013 & 0.070 & -0.099 & 0.188\textsuperscript{*} & 0.126 & 0.080 \\
    36.\ p\_see      & -0.012 & -0.010 & 0.090 & -0.115 & 0.169\textsuperscript{*} & 0.115 & 0.082 \\
    37.\ p\_relativ  & -0.212\textsuperscript{**} & -0.033 & 0.025 & -0.176\textsuperscript{*} & -0.061 & -0.149 & 0.032 \\
    38.\ p\_motion   & -0.021 & -0.011 & -0.053 & 0.040 & -0.071 & -0.081 & -0.100 \\
    39.\ p\_space    & -0.118 & -0.012 & 0.190\textsuperscript{*} & -0.235\textsuperscript{**} & 0.137 & 0.026 & -0.031 \\
    40.\ p\_time     & -0.099 & 0.040 & -0.144 & 0.013 & -0.157\textsuperscript{*} & -0.118 & 0.125 \\
    41.\ p\_work     & -0.009 & -0.062 & 0.354\textsuperscript{***} & -0.321\textsuperscript{***} & 0.052 & 0.029 & -0.295\textsuperscript{***} \\
    42.\ p\_achieve  & -0.132 & 0.058 & 0.016 & -0.168\textsuperscript{*} & 0.086 & 0.103 & -0.139 \\
    \bottomrule
  \end{tabular}
\end{table}
\addtolength{\tabcolsep}{3.5pt}

\addtolength{\tabcolsep}{-3.5pt}
\begin{table}[p]
  \centering
  \caption{Spearman's correlation of semantic variables in interactions in the direct condition (Part 3).}
  \label{tab:semVar.direct.3}
  \scriptsize
  \begin{tabular}{lccccccc}
    \toprule
    Variable & n\_see & n\_relativ & n\_motion & n\_space & n\_time & n\_work & n\_achieve \\
    \midrule
    22.\ p\_social   & -0.066 & -0.173\textsuperscript{*} & 0.037 & -0.219\textsuperscript{**} & 0.116 & -0.117 & -0.072 \\
    23.\ p\_affect   & 0.117 & -0.063 & 0.052 & -0.191\textsuperscript{*} & 0.162\textsuperscript{*} & -0.097 & -0.060 \\
    24.\ p\_posemo   & 0.107 & -0.067 & 0.055 & -0.192\textsuperscript{*} & 0.163\textsuperscript{*} & -0.101 & -0.060 \\
    25.\ p\_negemo   & 0.055 & 0.019 & -0.074 & 0.014 & -0.041 & 0.086 & 0.024 \\
    26.\ p\_cogmech  & -0.088 & -0.025 & 0.109 & 0.009 & -0.003 & -0.083 & -0.074 \\
    27.\ p\_insight  & -0.061 & -0.149 & 0.035 & -0.115 & 0.012 & -0.062 & -0.162\textsuperscript{*} \\
    28.\ p\_cause    & -0.068 & 0.160\textsuperscript{*} & 0.094 & 0.187\textsuperscript{*} & -0.013 & 0.020 & 0.125 \\
    29.\ p\_discrep  & -0.063 & -0.116 & -0.035 & 0.010 & -0.212\textsuperscript{**} & 0.059 & -0.177\textsuperscript{*} \\
    30.\ p\_tentat   & -0.082 & 0.038 & -0.103 & 0.116 & -0.064 & 0.076 & 0.086 \\
    31.\ p\_certain  & -0.202\textsuperscript{*} & -0.113 & -0.025 & -0.064 & -0.071 & 0.012 & 0.180\textsuperscript{*} \\
    32.\ p\_inhib    & -0.143 & -0.008 & 0.139 & -0.011 & -0.028 & -0.027 & 0.036 \\
    33.\ p\_incl     & -0.059 & -0.076 & -0.099 & -0.026 & -0.134 & 0.139 & -0.171\textsuperscript{*} \\
    34.\ p\_excl     & -0.127 & 0.041 & 0.004 & 0.042 & 0.066 & -0.028 & 0.119 \\
    35.\ p\_percept  & 0.073 & -0.177\textsuperscript{*} & -0.153 & -0.079 & -0.130 & 0.047 & -0.215\textsuperscript{**} \\
    36.\ p\_see      & 0.074 & -0.204\textsuperscript{*} & -0.164\textsuperscript{*} & -0.102 & -0.143 & 0.054 & -0.209\textsuperscript{**} \\
    37.\ p\_relativ  & 0.044 & 0.212\textsuperscript{**} & -0.027 & 0.213\textsuperscript{**} & -0.020 & 0.140 & 0.096 \\
    38.\ p\_motion   & -0.091 & 0.080 & 0.124 & 0.129 & -0.055 & -0.003 & -0.012 \\
    39.\ p\_space    & -0.029 & 0.072 & -0.190\textsuperscript{*} & 0.147 & -0.059 & 0.104 & 0.136 \\
    40.\ p\_time     & 0.114 & 0.114 & 0.106 & 0.012 & 0.062 & 0.052 & -0.042 \\
    41.\ p\_work     & -0.285\textsuperscript{***} & 0.023 & -0.196\textsuperscript{*} & 0.113 & -0.267\textsuperscript{***} & 0.564\textsuperscript{***} & 0.027 \\
    42.\ p\_achieve  & -0.173\textsuperscript{*} & -0.032 & -0.208\textsuperscript{**} & 0.049 & -0.124 & 0.231\textsuperscript{**} & -0.127 \\
    \bottomrule
  \end{tabular}
\end{table}
\addtolength{\tabcolsep}{3.5pt}

 \bibliographystyle{elsarticle-num} 
 \bibliography{navi}

\begin{thebibliography}{10}
\expandafter\ifx\csname url\endcsname\relax
  \def\url#1{\texttt{#1}}\fi
\expandafter\ifx\csname urlprefix\endcsname\relax\def\urlprefix{URL }\fi
\expandafter\ifx\csname href\endcsname\relax
  \def\href#1#2{#2} \def\path#1{#1}\fi

\bibitem{Chowa2025}
S.~S. Chowa, R.~Alvi, S.~S. Rahman, M.~A. Rahman, M.~A.~K. Raiaan, M.~R. Islam,
  M.~Hussain, S.~Azam, \href{https://arxiv.org/abs/2508.17281}{From language to
  action: A review of large language models as autonomous agents and tool
  users} (2025).
\newblock \href {https://doi.org/10.48550/ARXIV.2508.17281}
  {\path{doi:10.48550/ARXIV.2508.17281}}.
\newline\urlprefix\url{https://arxiv.org/abs/2508.17281}

\bibitem{Yi2025}
Z.~Yi, J.~Ouyang, Z.~Xu, Y.~Liu, T.~Liao, H.~Luo, Y.~Shen,
  \href{http://dx.doi.org/10.1145/3771090}{A survey on recent advances in
  llm-based multi-turn dialogue systems}, ACM Computing Surveys 58~(6) (2025)
  1–38.
\newblock \href {https://doi.org/10.1145/3771090} {\path{doi:10.1145/3771090}}.
\newline\urlprefix\url{http://dx.doi.org/10.1145/3771090}

\bibitem{Go2019}
E.~Go, S.~S. Sundar,
  \href{http://dx.doi.org/10.1016/j.chb.2019.01.020}{Humanizing chatbots: The
  effects of visual, identity and conversational cues on humanness
  perceptions}, Computers in Human Behavior 97 (2019) 304–316.
\newblock \href {https://doi.org/10.1016/j.chb.2019.01.020}
  {\path{doi:10.1016/j.chb.2019.01.020}}.
\newline\urlprefix\url{http://dx.doi.org/10.1016/j.chb.2019.01.020}

\bibitem{Feine2019}
J.~Feine, U.~Gnewuch, S.~Morana, A.~Maedche,
  \href{http://dx.doi.org/10.1016/j.ijhcs.2019.07.009}{A taxonomy of social
  cues for conversational agents}, International Journal of Human-Computer
  Studies 132 (2019) 138–161.
\newblock \href {https://doi.org/10.1016/j.ijhcs.2019.07.009}
  {\path{doi:10.1016/j.ijhcs.2019.07.009}}.
\newline\urlprefix\url{http://dx.doi.org/10.1016/j.ijhcs.2019.07.009}

\bibitem{cai2024communication}
N.~Cai, S.~Gao, J.~Yan, How the communication style of chatbots influences
  consumers’ satisfaction, trust, and engagement in the context of service
  failure, Humanities and Social Sciences Communications 11~(1) (2024) 1--11.

\bibitem{ding2024}
Y.~Ding, M.~Najaf, Interactivity, humanness, and trust: a psychological
  approach to ai chatbot adoption in e-commerce, BMC psychology 12~(1) (2024)
  595.

\bibitem{brown1987politeness}
P.~Brown, S.~C. Levinson, Politeness: Some universals in language usage,
  Vol.~4, Cambridge university press, 1987.

\bibitem{reeves1996media}
B.~Reeves, C.~Nass, The media equation: How people treat computers, television,
  and new media like real people, Cambridge, UK 10~(10) (1996) 19--36.

\bibitem{fiske2002model}
S.~T. Fiske, A.~J. Cuddy, P.~Glick, J.~Xu, A model of (often mixed) stereotype
  content: Competence and warmth respectively follow from perceived status and
  competition., Journal of Personality and Social Psychology 82~(6) (2002)
  878--902.
\newblock \href {https://doi.org/10.1037/0022-3514.82.6.878}
  {\path{doi:10.1037/0022-3514.82.6.878}}.

\bibitem{nass2000machines}
C.~Nass, Y.~Moon, Machines and mindlessness: Social responses to computers,
  Journal of social issues 56~(1) (2000) 81--103.

\bibitem{giles2023communication}
H.~Giles, A.~L. Edwards, J.~B. Walther, Communication accommodation theory:
  Past accomplishments, current trends, and future prospects, Language Sciences
  101 (2023) 101571.
\newblock \href {https://doi.org/10.1016/j.langsci.2023.101571}
  {\path{doi:10.1016/j.langsci.2023.101571}}.

\bibitem{giles2013communication}
H.~Giles, T.~Ogay, Communication accommodation theory, in: Explaining
  communication, Routledge, 2013, pp. 325--344.

\bibitem{Burgoon2005}
J.~K. Burgoon, A.~E. Hubbard, Cross-cultural and intercultural applications of
  expectancy violations theory and interaction adaptation theory, Theorizing
  about intercultural communication (2005) 149--171.

\bibitem{Sterken2024}
R.~K. Sterken, J.~R. Kirkpatrick,
  \href{http://dx.doi.org/10.1111/phpe.12205}{Conversational alignment with
  artificial intelligence in context}, Philosophical Perspectives 38~(1) (2024)
  89–102.
\newblock \href {https://doi.org/10.1111/phpe.12205}
  {\path{doi:10.1111/phpe.12205}}.
\newline\urlprefix\url{http://dx.doi.org/10.1111/phpe.12205}

\bibitem{liebrecht2020too}
C.~Liebrecht, L.~Sander, C.~Van~Hooijdonk, Too informal? how a chatbot’s
  communication style affects brand attitude and quality of interaction, in:
  International Workshop on Chatbot Research and Design, Springer, 2020, pp.
  16--31.

\bibitem{Svenningsson2019}
N.~Svenningsson, M.~Faraon,
  \href{http://dx.doi.org/10.1145/3375959.3375973}{Artificial intelligence in
  conversational agents: A study of factors related to perceived humanness in
  chatbots}, in: Proceedings of the 2019 2nd Artificial Intelligence and Cloud
  Computing Conference, AICCC 2019, ACM, 2019, p. 151–161.
\newblock \href {https://doi.org/10.1145/3375959.3375973}
  {\path{doi:10.1145/3375959.3375973}}.
\newline\urlprefix\url{http://dx.doi.org/10.1145/3375959.3375973}

\bibitem{bickmore2005establishing}
T.~W. Bickmore, R.~W. Picard, Establishing and maintaining long-term
  human-computer relationships, ACM Transactions on Computer-Human Interaction
  (TOCHI) 12~(2) (2005) 293--327.

\bibitem{islind2023friendly}
A.~S. Islind, M.~{\'O}skarsd{\'o}ttir, S.~{\'A}. Smith, E.~S. Arnard{\'o}ttir,
  \href{https://aisel.aisnet.org/scis2023/6}{The friendly chatbot: Revealing
  why people use chatbots through a study of user experience of conversational
  agents}, 14th Scandinavian Conference on Information Systems (2023).
\newline\urlprefix\url{https://aisel.aisnet.org/scis2023/6}

\bibitem{mairesse2011controlling}
F.~Mairesse, M.~Walker, Controlling user perceptions of linguistic style:
  Trainable generation of personality traits, Computational Linguistics 37~(3)
  (2011) 455--488.

\bibitem{pralat2024}
N.~Pralat, C.~Ischen, H.~Voorveld, Feeling understood by ai: How empathy shapes
  trust and influences patronage intentions in conversational ai, in:
  International Symposium on Chatbots and Human-Centered AI, Springer, 2024,
  pp. 234--259.

\bibitem{bhattacharjee2025}
D.~R. Bhattacharjee, A.~Kuanr, D.~Pradhan, T.~R. Moharana, Fun or warm: How
  conversational style boosts customer engagement, Journal of Retailing and
  Consumer Services 85 (2025) 104293.

\bibitem{deng2025}
Z.~Deng, J.~Yan, The effect of perceived warmth, competence, and social
  presence of ai-driven chabots on consumers’ engagement and satisfaction,
  SAGE Open 15~(3) (2025) 21582440251365438.

\bibitem{Fiske2007}
S.~T. Fiske, A.~J. Cuddy, P.~Glick,
  \href{http://dx.doi.org/10.1016/j.tics.2006.11.005}{Universal dimensions of
  social cognition: warmth and competence}, Trends in Cognitive Sciences 11~(2)
  (2007) 77–83.
\newblock \href {https://doi.org/10.1016/j.tics.2006.11.005}
  {\path{doi:10.1016/j.tics.2006.11.005}}.
\newline\urlprefix\url{http://dx.doi.org/10.1016/j.tics.2006.11.005}

\bibitem{Kulms2018}
P.~Kulms, S.~Kopp, \href{http://dx.doi.org/10.3389/fdigh.2018.00014}{A social
  cognition perspective on human–computer trust: The effect of perceived
  warmth and competence on trust in decision-making with computers}, Frontiers
  in Digital Humanities 5 (Jun. 2018).
\newblock \href {https://doi.org/10.3389/fdigh.2018.00014}
  {\path{doi:10.3389/fdigh.2018.00014}}.
\newline\urlprefix\url{http://dx.doi.org/10.3389/fdigh.2018.00014}

\bibitem{Lee2004}
J.~D. Lee, K.~A. See,
  \href{http://dx.doi.org/10.1518/hfes.46.1.50\_30392}{Trust in automation:
  Designing for appropriate reliance}, Human Factors: The Journal of the Human
  Factors and Ergonomics Society 46~(1) (2004) 50–80.
\newblock \href {https://doi.org/10.1518/hfes.46.1.50\_30392}
  {\path{doi:10.1518/hfes.46.1.50\_30392}}.
\newline\urlprefix\url{http://dx.doi.org/10.1518/hfes.46.1.50\_30392}

\bibitem{Gray2004}
J.~Gray, H.~Laidlaw,
  \href{http://dx.doi.org/10.1177/0893318903257980}{Improving the measurement
  of communication satisfaction}, Management Communication Quarterly 17~(3)
  (2004) 425–448.
\newblock \href {https://doi.org/10.1177/0893318903257980}
  {\path{doi:10.1177/0893318903257980}}.
\newline\urlprefix\url{http://dx.doi.org/10.1177/0893318903257980}

\bibitem{gnewuch2018faster}
U.~Gnewuch, S.~Morana, M.~Adam, A.~Maedche,
  \href{https://aisel.aisnet.org/ecis2018\_rp/113}{Faster is not always better:
  Understanding the effect of dynamic response delays in human-chatbot
  interaction}, in: Research Papers, 26th European Conference on Information
  Systems (ECIS 2018), 2018, p. 113.
\newline\urlprefix\url{https://aisel.aisnet.org/ecis2018\_rp/113}

\bibitem{Lee2011}
M.~K. Lee, S.~Kiesler, J.~Forlizzi,
  \href{http://dx.doi.org/10.1145/1978942.1978989}{Mining behavioral economics
  to design persuasive technology for healthy choices}, in: Proceedings of the
  SIGCHI Conference on Human Factors in Computing Systems, CHI ’11, ACM,
  2011, p. 325–334.
\newblock \href {https://doi.org/10.1145/1978942.1978989}
  {\path{doi:10.1145/1978942.1978989}}.
\newline\urlprefix\url{http://dx.doi.org/10.1145/1978942.1978989}

\bibitem{qin2025}
J.~Qin, Y.~Nan, Z.~Li, J.~Meng, Effectiveness of communication competence in ai
  conversational agents for health: Systematic review and meta-analysis,
  Journal of Medical Internet Research 27 (2025) e76296.

\bibitem{diederich2022}
S.~Diederich, A.~B. Brendel, S.~Morana, L.~Kolbe, On the design of and
  interaction with conversational agents: An organizing and assessing review of
  human-computer interaction research, Journal of the Association for
  Information Systems 23~(1) (2022) 96--138.

\bibitem{luger2016like}
E.~Luger, A.~Sellen, " like having a really bad pa" the gulf between user
  expectation and experience of conversational agents, in: Proceedings of the
  2016 CHI conference on human factors in computing systems, 2016, pp.
  5286--5297.

\bibitem{ruijten2015lonely}
P.~A. Ruijten, C.~J. Midden, J.~Ham, Lonely and susceptible: The influence of
  social exclusion and gender on persuasion by an artificial agent,
  International Journal of Human-Computer Interaction 31~(11) (2015) 832--842.

\bibitem{Nass1996}
C.~Nass, B.~Fogg, Y.~Moon, \href{http://dx.doi.org/10.1006/ijhc.1996.0073}{Can
  computers be teammates?}, International Journal of Human-Computer Studies
  45~(6) (1996) 669–678.
\newblock \href {https://doi.org/10.1006/ijhc.1996.0073}
  {\path{doi:10.1006/ijhc.1996.0073}}.
\newline\urlprefix\url{http://dx.doi.org/10.1006/ijhc.1996.0073}

\bibitem{tannen1990you}
D.~Tannen, You just don't understand: Women and men, Conversation. New York:
  Ballantine books (1990).

\bibitem{cross1997models}
S.~E. Cross, L.~Madson, Models of the self: self-construals and gender.,
  Psychological bulletin 122~(1) (1997) 5.

\bibitem{Sobieraj2020}
S.~Sobieraj, N.~C. Kr\"{a}mer,
  \href{http://dx.doi.org/10.1016/j.chb.2019.09.021}{Similarities and
  differences between genders in the usage of computer with different levels of
  technological complexity}, Computers in Human Behavior 104 (2020) 106145.
\newblock \href {https://doi.org/10.1016/j.chb.2019.09.021}
  {\path{doi:10.1016/j.chb.2019.09.021}}.
\newline\urlprefix\url{http://dx.doi.org/10.1016/j.chb.2019.09.021}

\bibitem{gefen1997gender}
D.~Gefen, D.~W. Straub, Gender differences in the perception and use of e-mail:
  An extension to the technology acceptance model, MIS quarterly (1997)
  389--400.

\bibitem{nazareth2019meta}
A.~Nazareth, X.~Huang, D.~Voyer, N.~Newcombe, A meta-analysis of sex
  differences in human navigation skills, Psychonomic bulletin \& review 26~(5)
  (2019) 1503--1528.

\bibitem{rheu2024}
M.~Rheu, Y.~Dai, J.~Meng, W.~Peng, When a chatbot disappoints you: Expectancy
  violation in human-chatbot interaction in a social support context,
  Communication Research 51~(7) (2024) 782--814.

\bibitem{bell2003}
L.~Bell, Linguistic adaptations in spoken human-computer dialogues-empirical
  studies of user behavior, Ph.D. thesis, Institutionen f{\"o}r
  tal{\"o}verf{\"o}ring och musikakustik (2003).

\bibitem{zellou2024}
G.~Zellou, N.~Holliday, Linguistic analysis of human-computer interaction,
  Frontiers in Computer Science 6 (2024) 1384252.

\bibitem{Araujo2024}
T.~Araujo, N.~Bol, \href{http://dx.doi.org/10.1016/j.chbah.2023.100030}{From
  speaking like a person to being personal: The effects of personalized,
  regular interactions with conversational agents}, Computers in Human
  Behavior: Artificial Humans 2~(1) (2024) 100030.
\newblock \href {https://doi.org/10.1016/j.chbah.2023.100030}
  {\path{doi:10.1016/j.chbah.2023.100030}}.
\newline\urlprefix\url{http://dx.doi.org/10.1016/j.chbah.2023.100030}

\bibitem{dahmani2023sex}
L.~Dahmani, M.~Idriss, K.~Konishi, G.~L. West, V.~D. Bohbot, Sex differences in
  spatial tasks: Considering environmental factors, navigation strategies, and
  age, Frontiers in Virtual Reality 4 (2023) 1166364.
\newblock \href {https://doi.org/10.3389/frvir.2023.1166364}
  {\path{doi:10.3389/frvir.2023.1166364}}.

\bibitem{Bruny2017}
T.~T. Brunyé, M.~D. Wood, L.~A. Houck, H.~A. Taylor,
  \href{http://dx.doi.org/10.1080/17470218.2016.1187637}{The path more
  travelled: Time pressure increases reliance on familiar route-based
  strategies during navigation}, Quarterly Journal of Experimental Psychology
  70~(8) (2017) 1439–1452.
\newblock \href {https://doi.org/10.1080/17470218.2016.1187637}
  {\path{doi:10.1080/17470218.2016.1187637}}.
\newline\urlprefix\url{http://dx.doi.org/10.1080/17470218.2016.1187637}

\bibitem{Wegmann2022}
A.~Wegmann, M.~Schraagen, D.~Nguyen,
  \href{https://aclanthology.org/2022.repl4nlp-1.26/}{Same author or just same
  topic? towards content-independent style representations}, in: S.~Gella,
  H.~He, B.~P. Majumder, B.~Can, E.~Giunchiglia, S.~Cahyawijaya, S.~Min,
  M.~Mozes, X.~L. Li, I.~Augenstein, A.~Rogers, K.~Cho, E.~Grefenstette,
  L.~Rimell, C.~Dyer (Eds.), Proceedings of the 7th Workshop on Representation
  Learning for NLP, Association for Computational Linguistics, Dublin, Ireland,
  2022, pp. 249--268.
\newblock \href {https://doi.org/10.18653/v1/2022.repl4nlp-1.26}
  {\path{doi:10.18653/v1/2022.repl4nlp-1.26}}.
\newline\urlprefix\url{https://aclanthology.org/2022.repl4nlp-1.26/}

\bibitem{mendes2023}
G.~A. Mendes, B.~Martins, Quantifying valence and arousal in text with
  multilingual pre-trained transformers, in: European Conference on Information
  Retrieval, Springer, 2023, pp. 84--100.

\bibitem{pennebaker1999linguistic}
J.~W. Pennebaker, L.~A. King, Linguistic styles: language use as an individual
  difference., Journal of personality and social psychology 77~(6) (1999) 1296.

\bibitem{hecht1978conceptualization}
M.~L. Hecht, The conceptualization and measurement of interpersonal
  communication satisfaction, Human Communication Research 4~(3) (1978)
  253--264.

\bibitem{lawton1994gender}
C.~A. Lawton, Gender differences in way-finding strategies: Relationship to
  spatial ability and spatial anxiety, Sex roles 30~(11) (1994) 765--779.

\bibitem{tausczik2010psychological}
Y.~R. Tausczik, J.~W. Pennebaker, The psychological meaning of words: Liwc and
  computerized text analysis methods, Journal of language and social psychology
  29~(1) (2010) 24--54.

\bibitem{benjamini1995controlling}
Y.~Benjamini, Y.~Hochberg, Controlling the false discovery rate: a practical
  and powerful approach to multiple testing, Journal of the Royal statistical
  society: series B (Methodological) 57~(1) (1995) 289--300.

\bibitem{coluccia2004gender}
E.~Coluccia, G.~Louse, Gender differences in spatial orientation: A review,
  Journal of environmental psychology 24~(3) (2004) 329--340.

\bibitem{schinazi2023motivation}
V.~R. Schinazi, D.~Meloni, J.~Grübel, D.~J. Angus, O.~Baumann, R.~P. Weibel,
  P.~Jeszenszky, C.~Hölscher, T.~Thrash, Motivation moderates gender
  differences in navigation performance, Scientific Reports 13~(15995) (2023).
\newblock \href {https://doi.org/10.1038/s41598-023-43241-4}
  {\path{doi:10.1038/s41598-023-43241-4}}.

\bibitem{brennan1990conversation}
S.~E. Brennan, Conversation as direct manipulation: An iconoclastic view, The
  art of human-computer interface design (1990) 393--404.

\bibitem{morozov2013save}
E.~Morozov, To save everything, click here: The folly of technological
  solutionism, PublicAffairs, 2013.

\bibitem{shneiderman2020human}
B.~Shneiderman, Human-centered artificial intelligence: Reliable, safe \&
  trustworthy, International Journal of Human--Computer Interaction 36~(6)
  (2020) 495--504.

\bibitem{Skitka2000}
L.~J. SKITKA, K.~MOSIER, M.~D. BURDICK,
  \href{http://dx.doi.org/10.1006/ijhc.1999.0349}{Accountability and automation
  bias}, International Journal of Human-Computer Studies 52~(4) (2000)
  701–717.
\newblock \href {https://doi.org/10.1006/ijhc.1999.0349}
  {\path{doi:10.1006/ijhc.1999.0349}}.
\newline\urlprefix\url{http://dx.doi.org/10.1006/ijhc.1999.0349}

\bibitem{Parasuraman2010}
R.~Parasuraman, D.~H. Manzey,
  \href{http://dx.doi.org/10.1177/0018720810376055}{Complacency and bias in
  human use of automation: An attentional integration}, Human Factors: The
  Journal of the Human Factors and Ergonomics Society 52~(3) (2010) 381–410.
\newblock \href {https://doi.org/10.1177/0018720810376055}
  {\path{doi:10.1177/0018720810376055}}.
\newline\urlprefix\url{http://dx.doi.org/10.1177/0018720810376055}

\bibitem{poivet2023influence}
R.~Poivet, M.~Lopez~Malet, C.~Pelachaud, M.~Auvray, The influence of
  conversational agents' role and communication style on user experience,
  Frontiers in Psychology 14 (2023) 1266186.
\newblock \href {https://doi.org/10.3389/fpsyg.2023.1266186}
  {\path{doi:10.3389/fpsyg.2023.1266186}}.

\bibitem{ryan2000self}
R.~M. Ryan, E.~L. Deci, Self-determination theory and the facilitation of
  intrinsic motivation, social development, and well-being., American
  psychologist 55~(1) (2000) 68.

\bibitem{dmello2017advanced}
S.~D'Mello, E.~Dieterle, A.~Duckworth, Advanced, analytic, automated (aaa)
  measurement of engagement during learning, Educational psychologist 52~(2)
  (2017) 104--123.

\bibitem{lakey2000social}
B.~Lakey, S.~Cohen, Social support theory and measurement, Social support
  measurement and intervention: A guide for health and social scientists 2952
  (2000).

\bibitem{ribino2023role}
P.~Ribino, The role of politeness in human--machine interactions: a systematic
  literature review and future perspectives, Artificial Intelligence Review
  56~(Suppl 1) (2023) 445--482.

\bibitem{tenenbaum2011telling}
H.~R. Tenenbaum, S.~Ford, B.~Alkhedairy, Telling stories: Gender differences in
  peers’ emotion talk and communication style, British Journal of
  Developmental Psychology 29~(4) (2011) 707--721.

\bibitem{hofstede1984culture}
G.~Hofstede, Culture's consequences: International differences in work-related
  values, Vol.~5, sage, 1984.

\end{thebibliography}

\end{document}